\documentclass[a4paper, 12 pt, english, reqno]{article}
\usepackage{titlesec}
\usepackage[margin=1.3in]{geometry}
\usepackage{setspace}
\usepackage[utf8]{inputenc}
\usepackage{xcolor}
\usepackage{bm}
\usepackage{amsmath}
\usepackage{amsthm}
\usepackage{amssymb}
\usepackage{bbm}
\usepackage[style=authoryear, uniquelist=false]{biblatex}

\usepackage{graphicx}
\usepackage[labelfont=bf, font=footnotesize, justification=justified, format=plain]{caption}
\usepackage{ragged2e}
\usepackage{subcaption}
\usepackage{float}
\usepackage{eurosym}
\graphicspath{{images/}}
\usepackage{booktabs}
\usepackage{xcolor}
\usepackage{cancel}
\usepackage{comment}

\usepackage{array}
\usepackage{lmodern,textcomp}
\newcolumntype{P}[1]{>{\centering\arraybackslash}p{#1}}
\newcommand{\beginsupplement}{
    \setcounter{table}{0}
    \renewcommand{\thetable}{A\arabic{table}}
    \setcounter{figure}{0}
    \renewcommand{\thefigure}{A\arabic{figure}}   
}
\usepackage{pdfpages}
\newtheoremstyle{theoremdd}
  {7mm}
  {7mm}
  {\itshape}
  {0pt}
  {\scshape}
  {. ---}
  { }
  {\thmname{#1}\thmnumber{ #2}\textnormal{\thmnote{ (#3)}}}
\theoremstyle{theoremdd}
\newtheorem{definition}{Definition}[section]
\newtheorem{lemma}{Lemma}[section]
\newtheorem{proposition}{Proposition}[section]
\usepackage{multicol}
\usepackage{multirow}
\renewcommand{\arraystretch}{1.5}

\usepackage{authblk}
\usepackage{hyperref}

\renewbibmacro*{cite}{%
  \printtext[bibhyperref]{%
    \iffieldundef{shorthand}{\ifthenelse{\ifnameundef{labelname}\OR\iffieldundef{labelyear}}
         {\usebibmacro{cite:label}%
          \setunit{\printdelim{nonameyeardelim}}}
         {\printnames{labelname}%
          \setunit{\printdelim{nameyeardelim}}}%
          \usebibmacro{cite:labeldate+extradate}}
      {\usebibmacro{cite:shorthand}}}}

\renewbibmacro*{citeyear}{%
  \printtext[bibhyperref]{%
    \iffieldundef{shorthand}
      {\iffieldundef{labelyear}
         {\usebibmacro{cite:label}}
         {\usebibmacro{cite:labeldate+extradate}}}
      {\usebibmacro{cite:shorthand}}}}

\renewbibmacro*{textcite}{%
  \ifnameundef{labelname}
    {\iffieldundef{shorthand}
       {\printtext[bibhyperref]{%
          \usebibmacro{cite:label}}%
        \setunit{%
          \global\booltrue{cbx:parens}%
          \printdelim{nonameyeardelim}\bibopenparen}%
        \ifnumequal{\value{citecount}}{1}
          {\usebibmacro{prenote}}
          {}%
        \printtext[bibhyperref]{\usebibmacro{cite:labeldate+extradate}}}
       {\printtext[bibhyperref]{\usebibmacro{cite:shorthand}}}}
    {\printtext[bibhyperref]{\printnames{labelname}}%
     \setunit{%
       \global\booltrue{cbx:parens}%
       \printdelim{nameyeardelim}\bibopenparen}%
     \ifnumequal{\value{citecount}}{1}
       {\usebibmacro{prenote}}
       {}%
     \usebibmacro{citeyear}}}

\renewbibmacro*{cite:shorthand}{%
  \printfield{shorthand}}

\renewbibmacro*{cite:label}{%
  \iffieldundef{label}
    {\printfield[citetitle]{labeltitle}}
    {\printfield{label}}}
\renewbibmacro*{cite:labeldate+extradate}{%
  \printlabeldateextra}

\definecolor{solblue}{HTML}{268BD2}
\hypersetup{
   pdfhighlight = /P,
   colorlinks=true,
   urlcolor=blue,
   linkcolor=purple,
   citecolor=solblue,
   raiselinks=true,
   breaklinks,
   unicode,
}

\usepackage[justification=centering]{caption} 
\usepackage[multiple]{footmisc}
\usepackage{placeins}

\usepackage{afterpage}
\usepackage{adjustbox}

\titleformat{\section}
  {\centering \scshape}
  {\thesection. } 
  {0pt}
  {}

\titleformat{\subsection}
  {\centering \itshape}
  {\thesubsection \hspace{1mm}} 
  {0pt} 
  {}

\titleformat{\paragraph}[runin]
{\slshape}{}{}{}

\title{\scshape The Network Effects of the EU Carbon Border Adjustment Mechanism with a Quantitative Trade Model\thanks{We are indebted to Lionel Fontagn\'e for initial discussion. We are also grateful to Mathieu Parenti, Maurizio Zanardi and Alessia Campolmi for useful comments.}}
\date{}
\author[1]{Noemi Walczak}
\author[2]{Kenan Huremović}
\author[3]{Armando Rungi}

\affil[1,2,3]{\small{\textit{IMT School for Advanced Studies Lucca, Lucca, Italy}}}
\affil[1]{\small \textit{Universit\'e Paris 1 Panthéon-Sorbonne, Paris, France}}

\begin{document}

\maketitle

\begin{abstract}

\vspace{-20mm}
\singlespacing
\noindent We investigate the economic and environmental impacts of the European Carbon Border Adjustment Mechanism (CBAM) using a multi-country, multi-sector general equilibrium model with input-output linkages adapted from \cite{caliendo2022distortions}. We quantify the general equilibrium responses of trade flows, expenditures, and emissions. To our knowledge, we are the first to endogenize both carbon prices and the CBAM price. We find that, when CBAM is at full force, it could contribute to reducing carbon emissions by $5.19\%$, a rate that could have been higher ($-8.84\%$) without the moderating role of global production networks. As a consequence of CBAM, we observe substitution effects towards non-targeted carbon-intensive inputs located upstream in the supply network. Notably, CBAM marginally increases EU Gross National Expenditure (GNE) due to changes in the terms of trade, and it shifts sourcing towards domestic and environmentally cleaner inputs. In contrast, extra-EU countries experience only a slight decline in GNE ($-0.02\%$) and emissions ($-0.11\%$). Finally, we propose counterfactual exercises that show the relative importance of technological progress on economic integration in global supply networks. Nonetheless, we argue that supply-chain-wise policies are useful for capturing the full carbon footprint in production.\\

    
\end{abstract}
\newpage
\vspace{5mm}
\section{Introduction}

Despite decades of warnings on climate change and the need for global policies, greenhouse gas (GHG) emissions have continued to rise throughout the 21st century \parencite{ipccreport}. Since 2005, a cornerstone of the European Union’s climate policy has been the Emission Trading System (ETS), also known as \textit{cap-and-trade}, a market-based mechanism that has proven effective in reducing emissions. While the ETS covers the EU as well as other countries in the European Economic Area agreement, its scope remains geographically limited. This creates a risk of carbon leakage, where emission-intensive production is relocated to countries with laxer climate regulations. To counter this, the EU introduced the Carbon Border Adjustment Mechanism (CBAM), which seeks to equalize the carbon cost between imported and domestically produced goods after the introduction of certificates whose price is linked to the weekly average price of the ETS. 

Since 2023, the CBAM has been in a transitional phase, requiring firms in selected polluting sectors to report the embodied emissions of their imports without financial obligations. Optimistically starting in 2026, importers would be required to purchase CBAM certificates corresponding to the embedded carbon content of covered goods. The scope is to reduce incentives to shift emissions abroad and strengthen the environmental integrity of the EU climate strategy\footnote{One of the criticisms made against the CBAM is that it would require a large information system to collect data about trade partners' emissions at a level of detail that makes it difficult to implement in a short time. See, for example, \cite{campolmi2024designing}.}.

Initially targeting six carbon-intensive goods, the mechanism is expected to expand to all sectors currently covered by the ETS by 2030. In this paper, we refer to the initial stage as the \textit{reduced CBAM}, and to the scenario where the CBAM is applied across all ETS sectors as the \textit{full CBAM}. As a cross-border policy instrument, the CBAM introduces a wedge between the cost of foreign producers and the price faced by importers, with potential implications for input sourcing, gains from trade, and broader macroeconomic outcomes.


Against this previous background, we adopt a quantitative trade model that is rich enough to account for both the complexity of the global supply network and the heterogeneity in climate policies across trade partner countries. In particular, we incorporate input-output linkages for each country-sector and endogenously determine the carbon prices in countries with a functioning emissions market. We calibrate the model using publicly available data collected from standard sources. We find that when the CBAM is fully implemented, it can lead to a significant reduction in emissions embodied in direct imports by 8.84\%. However, when we account for the emissions embodied in importers' supply chains, this number falls to 5.19\% due to substitution towards non-targeted inputs along the supply chains. When we focus on imported intermediates, we find that the CBAM can bring a decrease in the share of carbon-intensive (dirty) imported inputs by about 2.14\%, which is compensated by an increase in non-carbon-intensive (clean) imported and domestic inputs by 1.27\%, evidently suggesting a reallocation effect through changes in the map of relative costs of global inputs. Notably, similar reallocation effects can be found in the Gross National Expenditures of the new equilibrium, as we detect a small albeit positive increase (0.04\%). Ultimately, our findings appear to challenge the notion of a trade-off between eliminating carbon leakage, on the one hand, and gains from welfare and trade, on the other.


We start by building a multi-country, multi-sector general equilibrium model with input-output linkages, building on \textcite{caliendo2022distortions}. In the model, a representative firm from each sector in each country produces output using labor and the outputs of other sectors as inputs. Sectors can be located in the home country or in other countries. The production technology exhibits constant returns to scale and, importantly, generates emissions as a by-product. Markets are perfectly competitive, and trade in intermediate inputs at the country-sector level shapes the structure of global supply chains. In this context, we model the CBAM as a price levied on the carbon content of selected imported inputs. Importantly, this price is endogenously determined in the model: it is set to match the domestic price of the carbon content of the same sectoral goods, net of the foreign price abroad. To our knowledge, we are the first to endogenize carbon prices, and consequently the CBAM prices, with a quantitative trade model.

Intuitively, the introduction of CBAM increases the price of carbon-intensive imports from countries with laxer regulations, prompting a reallocation of demand towards cleaner foreign producers and EU-ETS producers. This shift influences the demand for ETS allowances, thereby affecting their market price and, through the endogenous mechanism, the CBAM price itself. Therefore, in the model, we capture not only the direct effects of the policy but also its feedback through emissions markets and allowance pricing. At the same time, our model is designed to capture the indirect effects emerging from the existence of complex production networks. An incentive like CBAM, which works as a wedge on direct imports, also has indirect effects upstream, i.e., when we consider, in turn, the importers' sourcing combination of clean and dirty inputs. 

We calibrate the model using publicly available and easily accessible data. In particular, we source data on bilateral trade flows, gross output, and value added from the OECD Inter-Country Input-Output tables, which collect information on 44 sectors across 32 countries, complemented by a residual category known as the Rest of the World. We obtain tariffs from the World Trade Organization (WTO) Integrated Database and Consolidated Tariff Schedule. Finally, we obtain the emissions data from the OECD Environmental Database, complementing it with the data from the European Union Transaction Log (EUTL) database. In countries with an operational national carbon pricing scheme by 2024, we set the total supply of emissions to match the aggregate quantities we retrieve from the respective databases mentioned above. For countries without a pricing scheme in place, we impose an exogenous carbon price equal to the country-specific Effective Carbon Rate, sourced from the OECD, while the supply of emissions is determined endogenously.


After our exercise, we begin by examining how the CBAM affects international trade patterns, focusing on the following three outcomes: the average share of intermediates the EU imports from abroad, the share of intermediates the EU sources domestically, and the size of EU sectors as measured by their average Domar weights in the world economy. Next, we discuss our core findings on the emissions embodied in EU imports and the issue of carbon leakage. We conclude by evaluating the impact of the CBAM on GNE in both the EU and extra-EU countries. For each outcome, we compare the effects of four different scenarios, derived from a $2 \times 2$ set of alternatives intended to provide a clearer picture of the policy effects. First, we examine the case of the \textit{reduced CBAM}, which focuses on a few polluting sectors, and the \textit{full CBAM}, which extends the policy to all sectors covered by the ETS. Thus, we distinguish the results of both the \textit{reduced CBAM} and \textit{full CBAM}, each with two additional alternative scenarios: (i) a benchmark case in which the CBAM value adjusts with emission prices, referred to as the \textit{endogenous} CBAM; and (ii) a case where the CBAM value is set according to the same mechanism but remains fixed, referred to as the \textit{exogenous} CBAM.

Let us consider the \textit{endogenous full CBAM} as our baseline, because it represents the original policy design by the EU authorities. In this case, we find that the CBAM leads to a roughly 2.14\% decline in the average share of imported carbon-intensive goods directly affected by the measure, encouraging a shift in demand toward cleaner alternatives. Accordingly, EU importers redirect demand towards non-carbon-intensive foreign and domestic input providers. Purchases of domestic goods grow by a positive albeit small coefficient (0.10\%), reflecting partial substitution between foreign intermediate goods and locally produced alternatives. The size of dirty EU sectors (measured using Domar weights) remains unchanged, while the size of clean EU sectors increases by 0.29\%, resulting from higher demand from both the EU and extra-EU countries. 


Our key insights relate to the changes in emissions embodied in imports. Let us focus more on this aspect. Our analysis distinguishes between direct emissions from imported goods and a broader measure that also includes indirect emissions embodied in the supply chains of EU importers. We find that the full application of CBAM reduces the emissions of direct imports by 8.84\%. When indirect imports are considered, the reduction falls to 5.19\%. Initially, the CBAM directly reduces imports of carbon-intensive goods that are explicitly targeted by the adjustment mechanism. However, this decline induces substitution towards non-targeted inputs, whose relative price falls and whose demand rises. As demand for these substitutes increases, so too does the demand for the inputs required to produce them -- including carbon-intensive ones. Importantly, reallocation weakens the overall impact of the CBAM because it increases carbon emissions upstream. The latter is an insight for the policymakers, who should consider the whole carbon footprint to evaluate the efficacy of the carbon policy \textit{versus} its targets. 

Interestingly, at the aggregate level, the full CBAM configuration increases real GNE by 0.04\%. The observed increase in EU welfare is primarily driven by terms-of-trade improvements resulting from a reallocation of demand that favors domestic production. These gains imply that the EU obtains more imports per unit of exports, thereby benefiting from improved trade conditions. In contrast, extra-EU countries experience a minor decline in real GNE with reductions of about 0.02\%.

Intuitively, the effects of CBAM strongly depend on two factors: the emissions intensity of production technologies in source countries and the degree of the EU integration into global supply chains. To better explore these dimensions, we conduct two types of counterfactual exercises. First, to isolate the role of technology, we vary a measure of production cleanliness in source countries. Second, we vary the degree of the EU integration into the global supply network. In both exercises, our primary outcome of interest is import-embodied emissions. Our results show that cleaner global production technologies reduce the impact of CBAM in a manner that is more than proportional. In contrast, deeper supply chain integration weakens the policy’s effectiveness by limiting the scope for substitution towards domestic inputs. Importantly, these two channels are interrelated: EU firms tend to source more heavily from countries that employ dirtier production technologies.

The remainder of the article is structured as follows. In Section \ref{sec: literature}, we relate to previous research. In Section \ref{sec: institutional framework}, we provide more details about the CBAM. In Section \ref{sec: model}, we introduce the theoretical framework, while Section \ref{sec:Quantification} shows how to calibrate and solve the model. Finally, Section \ref{sec:results} presents the results after the policy simulation and Section \ref{sec:conclusion} concludes and offers final thoughts on the future of CBAM.

\section{Related literature}\label{sec: literature}

This paper bridges two different fields of literature: the one on the trade effects of environmental policies, more specifically, the literature on carbon border adjustment, and the literature on production networks.

By extending the input-output model by \textcite{caliendo2022distortions} to incorporate emissions in production, we contribute to studying the interactions between international trade, carbon emissions, and environmental regulations thanks to a structural general equilibrium model. Previous studies include \textcite{duan2021environmental, larch2017carbon, larch2024consequences, korpar2023european, shi2025carbon}. Yet, unlike our framework, previous models do not map the global production network, either because they assume that producers choose to buy inputs only from the lowest-cost supplier -- as in the workhorse model by \textcite{caliendo2015estimates} -- or because they completely neglect the role of intermediate inputs in production.

In this respect, there is a strand of research on carbon border adjustments largely relying on traditional Computable General Equilibrium (CGE) models \parencite{bohringer2012alternative, ghosh2012border, bohringer2017targeted, morsdorf2022simple, bellora2023eu}. Unlike previous papers, our quantitative trade model tracks the economic mechanisms that generate the main results, thereby avoiding the rigidity and data intensity associated with CGEs.

The most relevant to our paper is the growing literature analyzing CBAM with structural trade models \parencite{sogalla2023unilateral, campolmi2024designing, florez2024eu, coster2024firms, farrokhi2025can}. In particular, the papers closest to ours are \textcite{florez2024eu} and \textcite{coster2024firms}. \textcite{florez2024eu} investigate the effect of the EU CBAM on trade and emissions with a general equilibrium model, while incorporating country-sectoral production linkages \textit{à la} \textcite{caliendo2015estimates}. They find a negative albeit small change in global emissions driven by the reallocation of production from carbon-intensive to less polluting countries. However, because their framework assumes that each sector imports a given input only from its lowest-cost supplier, it restricts the extent of reallocation across alternative input sources. By contrast, by mapping the full global input-output network, we quantify the propagation of CBAM through multiple direct and indirect channels, and we provide some insights on the country-sector-level heterogeneity. Our findings also complement those obtained by \textcite{coster2024firms}. They combine a structural model with product-level and balance sheet microdata, running quantitative simulations to compare carbon leakage and welfare effects of ETS-only \textit{versus} ETS-plus-CBAM scenarios. Their analysis focuses on French firms and captures firm-level sourcing decisions at both the intensive and extensive margin with monopolistic competition. They find that the CBAM reduces embedded emissions in French imports by inducing a reallocation of purchases towards cleaner inputs. By contrast, we abstract from firm heterogeneity and adopt a higher level of aggregation to study a global equilibrium after focusing on substitution effects along supply chains in the presence of heterogeneous sourcing strategies. 
Notably, while all previous papers treat the CBAM as an exogenous policy shock, we consider its endogenous nature as its specific design is such that it should interact with both EU and extra-EU climate policies, and favor the adoption of less carbon-intensive technologies.


Last but not least, our work also relates to literature on production networks \parencite{baqaee2020productivity, baqaee2024networks, carvalho2021supply}. In fact, while our model is fairly standard, we are the first to examine the network effects of a policy, the CBAM, which is crucial for solving international coordination problems in the presence of negative externalities, such as those induced by the use of polluting inputs. 

Overall, we argue that our framework provides a comprehensive assessment of the economic and environmental implications of CBAM, but also contributes to a broader discussion on the optimal design of policy tools in an interconnected global economy.

\section{Institutional setting}\label{sec: institutional framework}
Global greenhouse gas (GHG) emissions have increased steadily since the beginning of the 21st century, despite international efforts to curb them \parencite[]{ipccreport}. In response, the European Union has committed to reducing the net domestic greenhouse gas emissions by at least 55\% by 2030 relative to 1990 levels, and achieving climate neutrality by 2050 \parencite{FitFor55regulation}. Central to this strategy is the EU Emission Trading Scheme (ETS), the first major carbon pricing market, which also applies to Norway, Liechtenstein, Northern Ireland, and Iceland. The ETS is a \textit{cap-and-trade} mechanism: it sets a limit (cap) on a decreasing total amount of emissions that can be produced to achieve EU climate goals, then a corresponding amount of allowances, measured in tons of $\text{CO}_2$ equivalent, are distributed through a combination of auctioning and free allocation\footnote{Free allocation mainly targets sectors and sub-sectors that are considered most at risk of carbon leakage, in order to prevent the relocation of production to regions with less stringent climate policies. After the actual transitional phase, the share of free allowances should decrease while CBAM certificates become available.}.

The system covers around 40\% of European GHG emissions and, since its launch in 2005, it has helped drive down emissions from electricity, heat generation, and industrial production by 47\% \parencite{EUreport2023}\footnote{See Appendix Figure \ref{appendix_fig} for more details.}. 

\begin{figure}[h]
    \centering
    \includegraphics[width=0.9\linewidth]{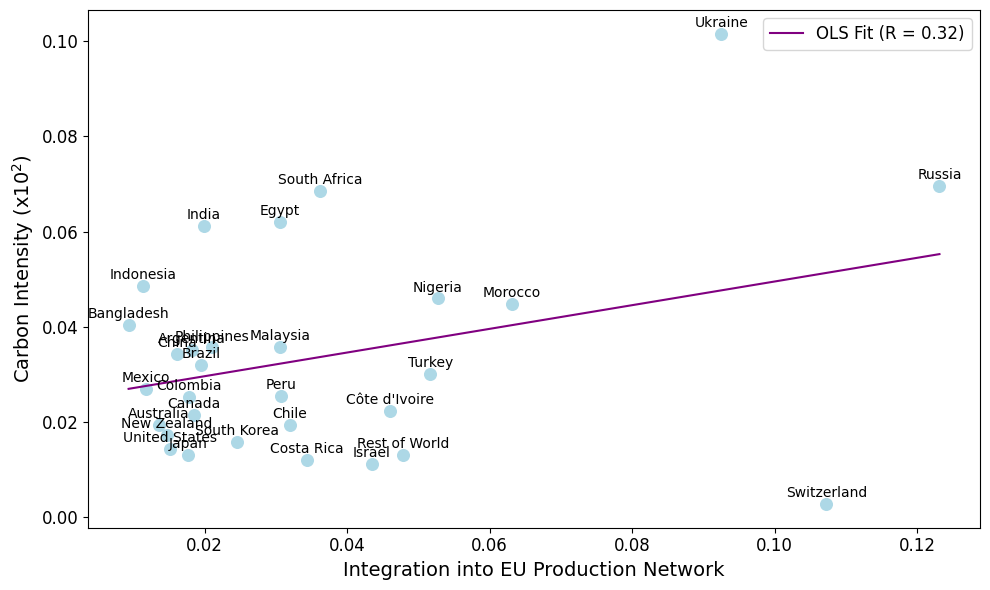}
    \caption{\justifying
    Carbon Intensity and Trade Integration into the EU supply network of non-ETS trading partners. Carbon Intensity is computed as the weighted average carbon intensity of each country, where the weights corresponds to the sector's share of the total country's sales. Trade Integration, instead, is computed as the country share of both direct and indirect exports to the EU, relative to the country's total sales.}
    \label{fig:motivation_CBAM}
\end{figure}

One consequence of the ETS is that producers covered by the system may face a competitive disadvantage compared to similar producers in countries without a comparable carbon pricing scheme. This creates an incentive for EU firms to relocate carbon-intensive stages of production to countries with less stringent environmental regulations—a phenomenon commonly referred to as \textit{carbon leakage}. As an illustration, Figure \ref{fig:motivation_CBAM} shows the positive correlation between each non-ETS country's weighted average carbon intensity and their integration into the European supply network. We calculate the weighted average carbon intensity by assigning a weight to each sectoral carbon intensity that corresponds to the sector's share of the total country's sales. Integration into the European supply network corresponds, instead, to the share of both direct and indirect exports to the EU, relative to the country's total sales.
The positive correlation underlines an important feature of the EU trade pattern: it is relatively highly integrated with countries that tend to have emission-intensive production structures. This pattern creates a misalignment between EU climate goals and the carbon content of its imports, undermining Europe's internal efforts by "offshoring" emissions through the purchase of carbon-intensive goods from foreign countries\footnote{After a second observation, Figure \ref{fig:motivation_CBAM} primarily underscores the heterogeneity in carbon intensity across EU partners. For more details, see also Appendix Table \ref{tab:rhos_appendix}}.  

To address these concerns, in 2021, the European Commission proposed the introduction of the Carbon Border Adjustment Mechanism (CBAM), complementing it with the gradual phase-out of the free allowances (Regulation (EU) 2023/956), which were previously provided to prevent moving production to countries with less strict carbon regulations. The CBAM is a policy tool designed to place a carbon price on the emissions embodied in imports of specific goods from non-EU countries, ensuring that importers face similar carbon costs as those borne by EU producers sourcing the same goods domestically. This aims to prevent carbon leakage and maintain the competitiveness of EU industries under the ETS.  Under the CBAM, European importers are required to purchase and submit CBAM certificates corresponding to the embedded emissions in their imported goods. The price of these certificates is based on the weekly average closing price of EU ETS allowances on the auction platform. This mechanism ensures that EU producers continue to face carbon costs under the EU ETS, while foreign exporters of emissions-intensive goods are virtually subject to a comparable carbon price, thereby leveling the playing field and reducing the risk of carbon leakage. By introducing a carbon price on the carbon content of imports, CBAM serves as a hybrid instrument of trade and climate policy, safeguarding the EU climate targets from being undermined by emissions embodied in imports.

Yet, two major downsides emerge. 
The first is that the design of a border adjustment mechanism, such as CBAM, discriminates across countries because the price of the same input varies depending on the origin country's level of emissions for that product. This is not compatible with the WTO clause on the Most Favored Nation (MFN), which implies that the same good imported from any two different origin countries must have the same tariffs or non-tariff barriers as in the most favored trade partner country. The point of view by the European Commission is that, once one evaluates the amount of regulatory burden, there is no discrimination between EU producers and any other partner country, and the mechanism is intended to equalize an otherwise unequal regulatory burden, therefore respecting both WTO principles of National Treatment (NT) and Most Favored Nation (MFN).

The second downside is that a well-functioning CBAM requires the establishment of an information system to collect and update data on emissions for different products in different countries. It may not be trivial to establish such a system in time for the end of the transitional phase. In fact, since 2023, the CBAM has been in a transitional phase, requiring firms in selected sectors to report the embodied emissions of their imports without financial obligations. Starting in 2026, importers will be required to purchase CBAM certificates corresponding to the carbon content of covered goods (iron and steel, cement, aluminum, organic basic chemicals, hydrogen, fertilizers, and electricity). The mechanism is expected to expand to all imported goods corresponding to those falling under the ETS scheme by 2030. By that time, a proper informative system is required to work at full capacity.


\section{The model} \label{sec: model}


We develop a multi-country, multi-sector model with country-sectoral linkages, building on \textcite{caliendo2022distortions}. There is a set $\mathcal{N} = \{ 1,..., N\}$ of countries, indexed by $i$ and $n$, and a set $\mathcal{J} = \{ 1,..., J\}$ of sectors for each country, indexed by $j$ and $k$. We use subscripts to denote countries, and superscripts to denote sectors. 

\subsection*{Intermediate and final goods production.}

Each sector $j$ in country $i$ is represented by a competitive producer that produces a single good using labor $l_i^j$ and materials $M_i^j$ as inputs. Each sectoral good can be either consumed as a final good by the household or used as an intermediate input in the production of other goods. Since we assume that each sector produces one good, we will use terms \textit{good/input} and \textit{sector} interchangeably when there is no fear of ambiguity. Following \textcite{copeland2003trade}, we assume that a fraction $a_i^j \in (0,1)$ of the input mix is used to abate pollution.
We use $q_{i}^j$  to denote the output of sector $j$ in country $i$, and assume that producers combine inputs in a Cobb-Douglas fashion. The production function exhibits constant returns to scale and has the following form:
\begin{equation}\label{eq:prod_fun_initial}
    q_i^j =  \Upsilon_i^j  A_i^j (1-a_i^j) (l_i^j)^{\beta_i^j} (M_i^j)^{1-\beta_i^j},
\end{equation}

\noindent where $A_i^j$ is a Hicks-neutral productivity parameter,  $l_i^j$ denotes the labor input, and $\beta_i^j > 0$ is its  share  in production.  $\Upsilon_i^j$ is a normalizing constant that we specify later. Moreover, we define materials $M_i^j$ as a CES aggregate of intermediate goods used in the production, namely:

\begin{equation}\label{eqCES}
    M_i^j = \left( \sum_{n \in \mathcal{N} }\sum_{k \in \mathcal{J}}(\iota_{ni}^{kj})^{\frac{1}{\theta}} (z_{ni}^{kj})^{\frac{\theta-1}{\theta}}  \right)^{\frac{\theta}{\theta-1}},
\end{equation}
 where $z_{ni}^{kj}$ is the amount of intermediate good produced by sector $k$ located in country $n$ used in the production of good $j$ in country $i$. Parameter $\theta$ governs the substitutability between intermediate inputs. The coefficient $\iota_{ni}^{kj} \ge 0$ measures the relevance that good of sector $k$ produced in country $n$ has in the production of good of sector $j$ in country $i$, with $ \sum_{n \in \mathcal{N}} \sum_{k \in \mathcal{J}} \iota_{ni}^{kj} = 1$. In particular, if $\iota_{ni}^{kj} = 0$, the specific good produced by sector $k$ in country $n$ is not used in the production of good $j$ in country $i$.

Crucially, we assume that each sector generates emissions $e_{i}^j$ as a by-product. The amount of emissions negatively depends on the abatement, with $\rho_i^j$ being the emissions elasticity:

\begin{equation}\label{eq:emissions}
    e_i^j = (1-a_i^j)^{\frac{1}{\rho_i^j}} \left[ A_i^j (l_i^j)^{\beta_i^j} (M_i^j)^{1-\beta_i^j} \right].
\end{equation}

Expressing $(1-a_i^j)$ from \eqref{eq:emissions} and plugging it into the production function \eqref{eq:prod_fun_initial}, total output $q_i^j$ can be rewritten as a function of both polluting emissions and inputs:
\begin{equation}\label{eqPF}
    q_i^j = \Upsilon_i^j \left[ A_i^j (l_i^j)^{\beta_i^j} (M_i^j)^{1-\beta_i^j} \right]^{1-\rho_i^j} [e_i^j]^{\rho_i^j}, 
\end{equation}
where, for convenience, we use the following normalization:
\begin{equation}\label{eq:normalization_upsilon}
\Upsilon_i^j
= \left[ (\beta_i^j)^{- \beta_i^j} (1-\beta_i^j)^{-(1-\beta_i^j)} \right]^{1-\rho_i^j}  (1-\rho_i^j)^{-(1-\rho_i^j)} (\rho_i^j)^{-\rho_i^j}.
\end{equation}

Given the functional form of the emissions' function, under the assumption that $a_i^j \in (0,1)$, there is a one-to-one mapping between the emissions $e_i^j$ and the abatement $a_i^j$. In the following, we treat $e_i^j$ as the producers' choice variable and use equation (\ref{eqPF}) as our production function.  The same approach is taken, for instance, in \textcite{shapiro2018pollution}.

\subsection*{Emissions.}

Following \textcite{bellora2023eu}, we distinguish between countries that have a national carbon market (belonging to the set $\mathcal{N}_c$) and countries that have not adopted any pricing mechanism on emissions ($\mathcal{N}_{nc}$). For the first group, we consider a national inelastic supply of emissions, and the unit price of emissions $t_i$ is determined endogenously to equate the aggregate demand for emissions with the available supply. This price can be interpreted as the shadow price of all regulations applying a price on emissions. Specifically, in \textit{cap-and-trade} systems, total national emissions are limited by the total number of allowances, each of which permits the emission of one unit $e_i^j$ of polluting emissions.
 Accordingly, $t_i$ represents the market price of these emission permits, which may either be purchased or allocated for free. We let $\epsilon_i^j \in (0,1)$ denote the share of emission permits that are freely distributed to a sector $j$ in country $i$, relative to the total number of permits, and adjust national supply by discounting for the free allowances. 
 In contrast, for countries in $\mathcal{N}_{nc}$, we impose an exogenous carbon price. In this case, the supply of emissions permits is endogenously determined to match the demand at the given price. 

\subsection*{International trade and CBAM.}

We assume that trade across countries and sectors is costly and subject to frictions modeled as a combination of iceberg trade costs $d_{ni}^{kj} \ge 1$, with $d_{ii}^{kj} = 1 \ \forall i,n, k,j$, and ad-valorem tariffs $\kappa_{ni}^{kj}$\footnote{We retain the destination sector index $j$ to capture potential heterogeneity in tariffs across importing sectors and to maintain flexibility for more realistic modeling. A similar approach is adopted, for example, in \textcite{baqaee2024networks}.}. Importantly, we directly incorporate the Carbon Border Adjustment Mechanism (CBAM) that targets the emissions embodied in trade, as an additional trade friction. Recall that the CBAM aims to ensure that imported goods face a comparable carbon cost to domestically produced goods, to mitigate carbon leakage and level the playing field. 

We model CBAM as a price wedge that equalizes the carbon cost of imported goods with the domestic carbon price, net of any carbon price already paid in the exporting country. When the exporting country imposes a higher, or equal, carbon price ($t_n \ge t_i$), we assume that no CBAM adjustment is applied.
The total price wedge for good $k$ originating in country $n$, when imported by sector $j$ in country $i$ is therefore given by:
\begin{equation}\label{eqCBAM}
    \tau_{ni}^{kj} =d_{ni}^{kj} \overbrace{(1+ \kappa_{ni}^{kj} + CBAM_{ni}^{kj})}^{\equiv \tilde{\tau}_{ni}^{kj}},
\end{equation}
\noindent where:
\begin{align}\label{eq:CBAM_definition}
CBAM_{ni}^{kj} = 
\begin{cases}
    \rho_n^k t_i \text{ if } t_n=0 \\
    \rho_n^k \frac{t_i}{t_n} \text{ if } t_i > t_n > 0, i \in \tilde{\mathcal{N}} \subseteq  \mathcal{\mathcal{N}}, n \notin \tilde{\mathcal{N}} \subseteq \mathcal{N}, k \in \tilde{\mathcal{J}} \subseteq \mathcal{\mathcal{J}}. \\
    0 \text{ otherwise }
\end{cases}
\end{align}
%
%
To simplify notation, we define $\tilde{\tau}_{ni}^{kj} \equiv 1 + \kappa_{ni}^{kj} + CBAM_{ni}^{kj}$. Our modeling approach is coherent with the predominant approach in literature that models carbon border adjustments as an additional tariff\footnote{Recent examples modeling the European CBAM as an additional tariff include  \textcite{bellora2023eu, campolmi2024designing, larch2024consequences, florez2024eu, coster2024firms}.}. 

We examine two distinct approaches to determining the value of CBAM, which we label \textit{exogenous} and \textit{endogenous} CBAM, respectively. When the value of CBAM is set based on the prevailing vector of emission prices $\mathbf{t} \in \mathbb{R}^{N}$, we refer to this as the exogenous case. However, because the introduction of CBAM generally alters the relative carbon prices across countries, we define the endogenous CBAM as the value that emerges when emission prices $\mathbf{t}$ adjust in response to the implementation of CBAM. The latter is a distinguishing feature of our approach that we haven't found in related work.

Therefore, under perfect competition, the price of one unit of good $k$ shipped from country $n$ to sector $j$ in country $i$ is given by:
\begin{equation*}
p_{ni}^{kj} = mc_n^k \tau_{ni}^{kj} =  \frac{1}{(A_n^k)^{1-\rho_n^k}} \left[ w_n^{\beta_n^k} (P_n^k)^{1-\beta_n^k}\right]^{1- \rho_n^k} \left[t_n(1-\epsilon_n^k) \right]^{\rho_n^k} \tau_{ni}^{kj}
\end{equation*}
 where $w_n$ is the wage in country $n$, $t_n$ is the price of carbon emissions, $\epsilon_n^k$ is the share of free allowances and $P_n^k = \left( \sum_{i \in \mathcal{N}} \sum_{j \in \mathcal{J}} \iota_{ni}^{kj}(p_{ni}^{kj})^{1-\theta} \right)^{ \frac{1}{1-\theta}} $ is the price index. 

\subsection*{Households.}

In each country $i$, there is a representative household supplying $\bar{L}_i$ units of labor inelastically. Moreover, in the subset of countries $\mathcal{N}_c \subseteq \mathcal{N}$ having a national carbon market, they also provide emissions permits, where $E_i$ denotes the total supply of emissions in country $i$. Preferences over sectoral final goods included in the bundle $\mathbf{c}_i = (c_i^1, ..., c_i^J)$ in each country $i \in \mathcal{N}$ are represented by the following CES utility function: 
\begin{equation}\label{eqHH}
    u(\mathbf{c}_i) = \left( \sum_{j \in \mathcal{J}} (\chi_i^j)^{\frac{1}{\sigma}} (c_i^j)^{\frac{\sigma -1}{\sigma}} \right)^{\frac{\sigma}{\sigma -1}} 
\end{equation}
 where $\chi_i^j \ge 0 $ denotes the weight that the consumption of good $j$ has in total consumption, with $\sum_{j \in \mathcal{J}} \chi_i^j = 1$ and $\sigma$ is the elasticity of substitution between final goods.

Households receive lump-sum transfers, derived from the collection of tariffs and CBAM, and transfers from the rest of the world in the form of exogenous trade deficits $\overline{D_i}$. The budget constraint is therefore given by:
$$ \sum_{j \in \mathcal{J}} p_i^j c_i^j \le I_i = w_i \overline{L}_i + t_i E_i +\sum_{j \in \mathcal{J}} \sum_{n \in \mathcal{N}} \sum_{k \in \mathcal{J}} (\kappa_{ni}^{kj}  + CBAM_{ni}^{kj}) p_{ni}^{kj}z_{ni}^{kj} + \overline{D_i} $$
where $\sum_{j \in \mathcal{J}} p_i^j c_i^j $ is country $i$'s Gross National Expenditure (GNE). Given locally non-satiated preferences, the constraint is satisfied with equality, and the solution to the consumer's problem yields the following optimal consumption of final good $c_i^j$:
$$ c_i^j = \frac{I_i}{\sum_{k \in \mathcal{J}} (\chi_i^k/ \chi_i^j) (p_i^k)^{1-\sigma}(p_i^j)^{\sigma}}  \quad \forall i \in \mathcal{N}, j \in \mathcal{J}$$
%

\subsection*{Equilibrium.}

We begin by defining the equilibrium in the case where the CBAM value is fixed based on the initial vector of emission prices $\mathbf{t}$ and does not respond to subsequent price changes triggered by the introduction of CBAM, as discussed earlier in this Section. We then extend this definition to allow the CBAM value to adjust endogenously in response to the price shifts induced by the trade barrier.
\begin{definition}[Equilibrium]\label{def:equilibrium}
    Given productivities $A_i^j$, wedges $d_{ni}^{kj}, \kappa_{ni}^{kj} $, the fixed values of  $CBAM_{ni}^{kj}$ and a vector of trade deficits $\overline{D_i}$ such that $\sum_{i \in \mathcal{N}} \overline{D_i} = 0$, the equilibrium is the set of wages $w_i^*$, carbon prices $t_i^*$ and prices of goods ${p_{i}^{j}}^*$, intermediate inputs choices ${z_{ni}^{kj}}^*$, factor input choices ${l_i^j}^*, {e_i^j}^*$, outputs ${q_i^j}^*$ and final demands ${c_i^j}^*$ such that: 
\begin{itemize}
    \item in each country, final demand maximizes the consumer's utility subject to the budget constraint;
    \item in each sector and country, producers maximize their profits, taking prices as given;
    \item markets for produced goods, labor, and emissions clear.
\end{itemize}
\end{definition}

In definition \ref{def:equilibrium}, the CBAM is determined according to equation \eqref{eq:CBAM_definition} with prices $\mathbf{t}$ equal to the equilibrium prices in the absence of CBAM --- that is, the equilibrium prices prior to the introduction of the CBAM. We refer to the resulting outcome as the equilibrium with \textit{exogenous} CBAM.
The equilibrium will be said to be the equilibrium with \textit{endogenous} CBAM when the prices $\mathbf{t}$ in \eqref{eq:CBAM_definition} are equal to the equilibrium prices for that level of CBAM. In the next section, we discuss how we solve for both types of equilibria.


\subsection*{Input-Output Definitions.}\label{sec:model_io_def}

Our model conceptualizes the world economy as a network, where each node represents a sector in a country, and the links denote the flows of intermediate inputs between these sector-country pairs. One of the goals of our analysis is to evaluate the role of the network's structure in mediating the effects of the CBAM. To ease the analysis of the model,  we introduce additional notation.

To facilitate calibration, we define the steady state of the economy as the situation in which wedges $d_{ni}^{kj}$ and $\kappa_{ni}^{kj}$ are equal to zero for all combinations of  $i$, $n$, $k$, $j$. Furthermore,  we normalize the Hicks productivity parameter $A_{i}^{j}$ such that ${(1-\epsilon_{i}^{j})}^{\rho_{i}^{j}}{A_{i}^{j}}^{-(1-\rho_{i}^{j})}=1$\footnote{Note that this is equivalent to appropriately rewriting the normalization constant $\Upsilon_{i}^{j}$ in \eqref{eq:normalization_upsilon}.}. 

\begin{itemize}
    \item The \textit{cost share} of intermediate input $k$ produced in country $n$ in the total intermediate inputs used in sector $j$ in country $i$ is denoted by $\tilde{\omega}_{ni}^{kj}$ and, in equilibrium, it corresponds to: 

    $$\tilde{\omega}_{ni}^{kj} \equiv \frac{p_{ni}^{kj} z_{ni}^{kj}}{P_i^j M_i^j} = \iota_{ni}^{kj}  \left( \frac{p_{ni}^{kj}}{P_i^j} \right)^{1-\theta} = \frac{\iota_{ni}^{kj} (p_{ni}^{kj})^{1-\theta}} {\sum_{n \in \mathcal{N}} \sum_{k \in \mathcal{J}} \iota_{ni}^{kj} (p_{ni}^{kj})^{1-\theta} }$$

    We denote the set of $N \times M$ real matrices by $\mathcal{M}(N \times M)$ and let $\mathbf{\tilde{\Omega}} \in \mathcal{M}(NJ \times NJ)$ be the matrix with elements $\tilde{\omega}_{ni}^{kj}$, which records the direct inter-country and inter-sector flows of intermediate goods. In the steady state this matrix coincides with $\mathbf {\Pi} \in \mathcal{M}(NJ \times NJ)$, with entries $\iota_{ni}^{kj}$. 

    \item The expenditure share  of intermediate input $k$ produced in country $n$ in the total sales of sector $j$ in country $i$, is denoted by $\omega_{ni}^{kj}$ and, in equilibrium, it corresponds to the following \textit{revenue share}: 

    $$\omega_{ni}^{kj} \equiv \frac{1}{\tilde{\tau}_{ni}^{kj}} \frac{p_{ni}^{kj}z_{ni}^{kj}}{p_i^jq_i^j} =(1-\beta_i^j)(1-\rho_i^j)  \frac{\tilde{\omega}_{ni}^{kj}}{\tilde{\tau}_{ni}^{kj}} $$

    Similarly, let $\mathbf{\Omega} \in \mathcal{M}(NJ \times NJ)$ be the matrix with entries $\omega_{ni}^{kj} $ and define the Leontief inverse $\mathbf{\Psi} \in \mathcal{M}(NJ \times NJ)$ with entries $\psi_{ni}^{kj}$ as:
    
    $$\mathbf{\Psi} \equiv [\mathbf{I} - \mathbf{\Omega}]^{-1} = \mathbf{I} + \mathbf{\Omega} + \mathbf{\Omega}^2 + ...$$
    
    The matrix $\mathbf{\Psi}$ accounts for all direct and indirect linkages of the production network and, in the steady state, it corresponds to $\bm{\Psi} = (\mathbf{I}- \bm{\gamma} \mathbf{\Pi}')^{-1}$, where $\bm{\gamma} = diag(\gamma_1^1, ...,\gamma_N^J) \in \mathcal{M}(NJ \times NJ) $ and $\gamma_i^j = (1-\beta_i^j)(1-\rho_i^j)$.

    \item The share of good $j$ purchased by the representative household in country $i$ is denoted by $\alpha_i^j$ and, in equilibrium, it corresponds to the following \textit{consumption share}:

    $$ \alpha_i^j \equiv \frac{p_i^j c_i^j}{\sum_{j \in \mathcal
    J} p_i^j c_i^j} = \frac{p_i^j c_i^j}{I_i} = \frac{ \chi_i^j (p_i^j)^{1-\sigma}}{ \sum_{j \in \mathcal{J}} \chi_i^j (p_i^j)^{1-\sigma}}$$

    In the steady state, it coincides with $\chi_i^j$.

    \item Total sales as a share of nominal world GNE are equal to the \textit{Domar weights} (in the world economy) and they are defined as:

    $$\lambda_{i}^{j} \equiv \frac{p_i^jq_i^j}{GNE} = \frac{p_i^jq_i^j}{\sum_{i \in \mathcal{N}} \sum_{j \in \mathcal{J}} p_i^j c_i^j}$$
    
\end{itemize}

\section{Quantifying the effect of CBAM} \label{sec:Quantification}

In this section, we apply the model developed in the previous paragraphs to evaluate the effects of introducing the CBAM on trade across sectors and countries, emissions embodied in imports, carbon leakage, gross national expenditure, and real wages.

First, we solve the model in changes using the exact hat-algebra approach \parencite{dekle2008global}, which enables us to evaluate different counterfactual scenarios. In the following exercises, we consider both adjustments: the full non-linear adjustments to a finite — though not necessarily marginal — policy shock, while maintaining computational efficiency through a reduced number of parameters. To illustrate the mechanisms at play, we provide a comparative static analysis of a marginal adjustment in CBAM in the Appendix \ref{appendix_proofs}.

In our following exercises, we consider both an adjustment to the equilibrium with an exogenous CBAM and an endogenous CBAM, as explained in the previous section. That is, unlike standard fixed tariffs, the CBAM adjusts in response to domestic and foreign carbon prices -- both of which may be influenced by the implementation of CBAM itself. This mechanism is intended to provide an incentive for faster convergence in clean technologies and regulations. 

By considering both scenarios, we can assess the significance of the additional adjustment mechanism in both the case of a reduced CBAM, i.e., when the policy applies only to a subset of carbon-intensive goods, and the case of a full CBAM, i.e., when the policy applies to all sectors of the EU economy currently covered by the ETS. Throughout the remainder of the analysis, we assume that the reduced CBAM applies to the following six sectors in our data: Mining and Quarrying; Chemicals and Chemical Products; Non-Metallic Mineral Products; Basic Metals; Fabricated Metal Products; and Electricity, Gas, Steam, and Air Conditioning Supply. 

\subsection*{Solving the model in changes.}  We use the exact hat-algebra following \textcite{dekle2008global} to characterize a counterfactual equilibrium in terms of proportional changes relative to the steady state.
Specifically, the equilibrium in relative changes is computed by solving the following system of nonlinear equations that jointly determine changes in wages, intermediate input prices, carbon prices, and emissions, as well as cost and consumption shares, and expenditures.

\begin{definition}
    For any variable, let $x$ denote the value before the introduction of the CBAM and $x'$ denote the counterfactual value. Define $\hat{x} \equiv \frac{x'}{x}$ as the relative change of the variable. The equilibrium conditions in relative changes satisfy:
    
    \begin{itemize}
        \item Cost of the input bundle:        
        \begin{equation}\label{eq:cost_changes}
            \hat{mc}_i^j = \left(\hat{w}_i^{\beta_i^j} {(\hat{P}_i^j)} ^{1-\beta_i^{j}} \right)^{1-\rho_i^j}\left( \hat{t}_i (1-\hat{\epsilon}_i^j) \right)^{\rho_i^j}
        \end{equation}

        \item Tariffs:

        \begin{equation}\label{eq:tariff_changes}
            \hat{\tau}_{ni}^{kj'} = \frac{1 + \kappa_{ni}^{kj} + \rho_n^k\frac{t'_i}{t'_n} \nu_{i,n} }{1 + \kappa_{ni}^{kj} }
        \end{equation}

        where $\nu_{i,n}$ is the ratio between the observed carbon prices of country $i$ and $ n$\footnote{In our model, quantities are normalized using specific country-sector parameters. This implies that the equilibrium carbon prices may correspond to different physical amounts of emissions. In order to compare them in a meaningful way, we need to map the equilibrium carbon prices back to the observed carbon prices, which we do through the constant $\nu_{i,n}$.}.

        \item Price index: 

        \begin{equation}\label{eq:price_changes}
          \hat{P}_i^j = \left( \sum_{n \in \mathcal{N}} \sum_{k \in \mathcal{J}}{\iota}_{ni}^{kj} (\hat{mc}_n^{k} \hat{\tau}_{ni}^{kj})^{1-\theta} \right)^{\frac{1}{1-\theta}}  
        \end{equation}
        
        \item Cost shares:

        \begin{equation}\label{eq:cost_shares_changes}
            \hat{\tilde{\omega}}_{ni}^{kj} = \left( \frac{\hat{mc}_n^{k}\hat{\tau}_{ni}^{kj}} {\hat{P}_i^j} \right)^{1-\theta}
        \end{equation}

        \item Consumption shares: 

        \begin{equation}\label{eq:consumption_shares_changes}
            \hat{\alpha}_i^j = \left( \frac{\hat{p}_i^j}{ \left( \sum_{j \in \mathcal{J}} \chi_i^j (\hat{p}_i^j)^{1-\sigma} \right)^{\frac{1}{1-\sigma} }}  \right)^{1-\sigma}  
        \end{equation}

        where $\hat{p}_i^j = \hat{mc}_i^j$.

        \item Total sales of in each country i and sector j:

        \begin{equation}\label{eq:expenditure_changes}
            p_i^{j'} q_i^{j'} =  \alpha_i^{j'} I'_i + \sum_{n \in \mathcal{N}} \sum_{k \in  \mathcal{J}} \omega_{in}^{jk'} p_n^{k'} q_n^{k'}
        \end{equation}

        where $I'_i = w_i'\overline{L_i} + t'_i E'_i + \sum_{j \in \mathcal{J}} p_i^{j'} q_i^{j'} \sum_{n \in \mathcal{N}} \sum_{k \in \mathcal{J}} (\tilde{\tau}_{ni}^{kj'} -1)\omega_{ni}^{kj} + \overline{D_i}$

        \item Labor market clearing:

        \begin{equation}\label{eq:wage_changes}
            \hat{w}_i = \frac{1}{w_i\overline{L_i}}\sum_{j \in \mathcal{J}} \beta_i^j(1-\rho_i^j) p_i^{j'} q_i^{j'}  
        \end{equation}

        \item Emission market clearing:

        \begin{equation}\label{eq:emission_changes}
            \begin{cases}
                \hat{t}_i =  \frac{1}{t_i\overline{E_i}} \sum_{j \in \mathcal{J}} \rho_i^j p_i^{j'} q_i^{j'} \quad \forall i \in \mathcal{N}_c \\ 

                \hat{E}_i = \frac{1}{\overline{t_i} E_i } \sum_{j \in \mathcal{J}} \rho_i^j p_i^{j'} q_i^{j'} \quad \forall i \in \mathcal{N}_{nc} 
            \end{cases}
        \end{equation}
    \end{itemize}
\end{definition}

Please note that the endogenous nature of CBAM is visible from equation \eqref{eq:tariff_changes}, where the part of the total tariffs which is due to CBAM ($\rho_n^k \frac{t'_i}{t'_n}$) depends on the counterfactual values of prices. In the case of exogenous CBAM, this term changes to ($\rho_n^k \frac{t_i}{t_n}$), where $t_i$ and $t_n$ are the price levels in the equilibrium prior to the introduction of CBAM. 

We solve the system using a fixed-point iteration algorithm, similar to the one used in \textcite{caliendo2015estimates, mayer2019cost}, following the steps below:

\begin{enumerate}
    \item Guess the initial vectors of changes for $\mathbf{w}$, $\mathbf{t} \in \mathbb{R}^N$;
    \item Use equation (\ref{eq:tariff_changes}) to derive the change in tariffs, after the introduction of the CBAM in the counterfactual scenario;
    \item Use Equations (\ref{eq:price_changes})-(\ref{eq:consumption_shares_changes}) to solve for the change in prices, cost shares and consumption shares;
    \item Given the change in trade shares and income, retrieve the counterfactual level of expenditure for each country-sector pair through equation (\ref{eq:expenditure_changes}) ;
    \item Aggregating over sectors, labor market clearing in (\ref{eq:wage_changes}) and market clearing for emissions in (\ref{eq:emission_changes}) imply an update of the initial conditions, dampened by a factor of 0.1.
    \item Iterate until convergence, i.e., until the norm of the difference between successive iterations falls below a given threshold.
\end{enumerate}

\subsection*{Calibration and the baseline.} 

Solving the model in relative changes has the advantage of reducing the data needed for the calibration. We rely on a minimal yet sufficient set of macroeconomic and environmental data, which enhances the transparency, tractability, and ease of replication of our exercise. Specifically, we calibrate the model using data from the last available year, 2018 (see \ref{appendix_data} for a detailed exposition of the model's mapping to the data). Following \textcite{bellora2023eu}, we construct the baseline year (2024) by incorporating environmental policies implemented between 2018 and 2024. This is important to account for the reduction in energy-related $\text{CO}_2$ emissions observed in advanced economies during this period. The effects of the CBAM are then assessed relative to the baseline year 2024. Further details are provided later in this section. 

In particular, we work on bilateral trade flows in intermediate inputs and final goods, gross output and value added sourced from the OECD Inter-Country Input-Output tables \parencite[]{oecd2021}, which collect data from 44 sectors across 32 countries, complemented by a residual Rest of the World to close the global matrix. With these data, we can retrieve the empirical counterparts for the steady state values of the cost shares $\iota_{ni}^{kj}$, the labor input shares $\beta_i^j$, the consumption shares $\chi_i^j$, and the exogenous deficits $\bar{D_i}$. Then, we source tariffs from the WTO Integrated Database and Consolidated Tariff Schedule\footnote{We follow \textcite{baqaee2024networks} and calibrate the model using observed trade value data and emissions, adjusting for existing tariffs. This practically means that we set $\kappa_{ni}^{kj} =0$ at the baseline in our quantitative exercise.}, while we retrieve Scope 1 emissions from production from the OECD Environmental Statistics database. In order to build the emissions' elasticity $\rho_i^j$, emissions quantities are multiplied by the corresponding country’s Effective Carbon Rate \parencite{ECR2023}, converted into USD using the OECD exchange rates. For the European Union, Iceland, and Norway, we employ a different approach. We collect emissions data from the European Union Transaction Log (EUTL) database. We aggregate them at the selected sectoral level and adjust the reported emissions by deducting those covered by free allowances granted to the installations operating in Energy Intensive Trade Exposed (EITE) industries. Note that this choice implies that we will evaluate lower bounds for both the level and the variation of emissions in these countries, as the data source excludes emissions from entities not covered by the ETS. For this reason, we refer to \textit{EU emissions} and \textit{ETS emissions} interchangeably, unless the distinction is necessary to avoid ambiguity.  

In countries with an operational national carbon pricing scheme by 2024, we set the total supply of emissions to match the aggregate quantities we retrieve from the databases described above. For countries without a pricing scheme in place by 2024\footnote{We collect details about direct carbon pricing initiatives around the world on the \textit{World Bank Carbon Pricing Dashboard} (\url{https://carbonpricingdashboard.worldbank.org/}). The full list includes: Bangladesh, Brazil, Côte d'Ivoire, Costa Rica, Egypt, India, Israel, Morocco, Malaysia, Nigeria, Peru, Philippines, Russia, Turkey, and the United States. Moreover, we include China in this list, since we believe its current environmental regulations are not stringent enough to incentivize decarbonisation efforts.}, we impose an exogenous carbon price equal to the country-specific Effective Carbon Rate, while the supply of emissions is determined endogenously. Finally, consistent with \textcite{caliendo2022distortions}, we assume an elasticity of substitution between intermediate goods and final consumption goods equal to 4, constant across countries and sectors.

\paragraph{The construction of the baseline.}

Before introducing the CBAM, we run a simulation in which, following \textcite{bellora2023eu}, selected countries implement their emissions reduction pledges under the Paris Agreement. Then, we use the resulting equilibrium as our baseline. Rather than comparing the scenario where the CBAM is introduced with the original 2018 steady state, we adopt this policy-adjusted baseline for two main reasons. First, the implementation of the Paris Agreement commitments is binding for many countries, making it more representative of the environment where the CBAM operates. Second, it enables us to isolate the specific contribution of the CBAM, net of ongoing climate policy changes and the related adjustments in emissions observed in recent years.

For the baseline, we select the countries with a carbon pricing scheme and consider the unconditional targets declared in their 2020 Nationally Determined Contributions (NDCs) under the Paris Agreement\footnote{We consider the most recent unconditional pledges submitted to the NDCs Registry as of 2024. Conditional targets are excluded from this exercise.}. We convert all the considered commitments into a fixed target level for the year 2030 and derive a yearly percentage reduction rate from the declared base year, consistent with each country's target. We then apply an exponential decay from 2018 to 2024 to model emission trajectories. Two notable exceptions apply. First, regarding ETS countries, we also consider a 40\% reduction in the freely allocated allowances relative to 2018 levels, as supported by empirical data (see Figure \ref{figETSemissions}). Second, as for China, we follow \textcite{bellora2023eu} in assuming the absence of a fully operational carbon market. We believe that its current carbon price is too low to reach the targets declared in the country's NDC\footnote{According to the \textcite{ECR2023}, in the last year recorded, coinciding with 2023, the Effective Carbon Rate was EUR 7.27 per tonne of $\text{CO}_2\text{e}$.} and, accordingly, we treat China's carbon price as exogenous. 

We impute the computed change in emissions between 2018 and 2024 and simulate the counterfactual equilibrium. The resulting cost shares and consumption shares, together with the new pollution intensity parameters, are then used to recalibrate the model and build our new baseline.

\paragraph{Policy scenarios.} 
We assume that standard trade tariffs remain constant and introduce the CBAM as a price wedge applied to the embodied emissions in imports of targeted goods entering the EU, as defined in equation (\ref{eqCBAM}). Given the gradual phase-in of the policy specified in Regulation (EU) 2023/956, we consider two counterfactual scenarios:

\begin{itemize}
    \item \textit{Reduced CBAM}: under the Regulation, the border adjustment initially applies only to specific goods -- iron and steel, cement, aluminum, organic basic chemicals, hydrogen, fertilizers, and electricity. In our exercise, they correspond to six broader sectors: Mining and Quarrying, Chemicals and Chemical Products, Non-Metallic Mineral Products, Basic Metals, Fabricated Metal Products, and Electricity, Gas, Steam, and Air Conditioning Supply;
    
    \item \textit{Full CBAM}: in this scenario, the adjustment is extended to all imported goods corresponding to the categories falling under the EU-ETS coverage. 
\end{itemize}

\section{Results}\label{sec:results}

In the following paragraphs, we present our main findings. We begin by analyzing how the CBAM affects international trade patterns, focusing on the following three outcomes: the average share of intermediate inputs the EU imports from abroad, the share of intermediate inputs the EU sources domestically, and the size of EU sectors as measured by their average Domar weights in the world economy. We report these results in Table \ref{tableIMPEXP}. Next, we turn to our core findings on carbon leakage and emissions embodied in EU imports, summarized in Table \ref{tableEEI}. We continue by evaluating the impact of the CBAM on welfare gains, as proxied by gross national expenditure (GNE) in both the EU and extra-EU countries in Table \ref{tableGDPEMISSIONS}. For each outcome, we compare the implications in our $2 \times 2$ scenarios. On the one hand, we present the reduced \textit{versus} full CBAM and, on the other hand, we distinguish between scenarios in which the value of the CBAM is set either endogenously or exogenously, as discussed in Section \ref{sec: model}. 

Our preferred specification is the one with full and endogenous CBAM, which corresponds to the scenario in which the original policy design is fully in effect. When discussing our results in the following paragraphs, we will use the terms \textit{dirty}/\textit{carbon-intensive} goods to refer to the sectoral goods falling under the ETS scheme\footnote{See Appendix Table \ref{tableA1} for the classification of sectors falling under the ETS and/or the CBAM.}, while classifying the residual sectoral goods as \textit{clean}/\textit{non-carbon-intensive}.

Finally, we conclude our contribution with two counterfactual exercises to separate the roles of international trade and technology in Figure \ref{fig:counterfactuals_rhos}. \textit{Ceteris paribus}, we first vary the output elasticity of emissions, then the share of imported inputs.

\subsection{Trade patterns.}\label{sec:trade patterns}
Let us start with trade patterns. By design, the CBAM immediately increases the relative price of imported carbon-intensive goods at the EU border, by creating a price wedge that worsens the relative competitiveness of foreign producers. Therefore, we first check to what extent CBAM causes substitution away from dirty imported inputs. In panel (a) of Table \ref{tableIMPEXP}, this mechanism leads to a decline in the average share of foreign purchases, which is the average share of European imports from all other countries in total intermediate input purchases. In case of a reduced CBAM, the change in the foreign purchases share is $-0.6\%$, while in the case of the full CBAM it more than doubles to $-1.27\%$. 

\begin{table}[p]
\centering
\scalebox{0.75}{
\begin{tabular}{l c c c c c c c c}
\toprule
\multirow{3}{*}{\textbf{Variable}} 
  & \multicolumn{4}{c}{\textbf{Reduced CBAM}} 
  & \multicolumn{4}{c}{\textbf{Full CBAM}} \\
  \cmidrule(rl){2-5} \cmidrule(rl){6-9}
 & \multicolumn{2}{c}{\textbf{\textit{Endog.}}} &  \multicolumn{2}{c}{\textit{Exog.}} &  \multicolumn{2}{c}{\textbf{\textit{Endog.}}}&  \multicolumn{2}{c}{\textit{Exog.}}  \\
 \cmidrule(rl){2-3} \cmidrule(rl){4-5}  \cmidrule(rl){6-7} \cmidrule(rl){8-9}
 & Mean & St.dev. & Mean & St.dev. & Mean & St.dev. & Mean & St.dev. \\

\midrule
\multicolumn{9}{c}{\textbf{Panel (a)} $\Delta \%$ in EU Share of Foreign Purchases} \\
\textit{Total} 
  & \textbf{-0.60} & \textbf{1.30} & -0.57 & 1.25 
  & \textbf{-1.27} & \textbf{2.80} & -1.23 & 2.73 \\
\quad \dots \textit{of clean intermediate goods} 
  & \textbf{0.57} & \textbf{0.34} & 0.55 & 0.33 & \textbf{1.91} & \textbf{1.70} & 1.86 & 1.66 \\
\quad \dots \textit{of dirty intermediate goods} 
  & \textbf{-1.06} & \textbf{2.37} & -1.02 & 2.28  
  & \textbf{-2.14} & \textbf{3.59} & -2.07 & 3.49 \\
\midrule
\multicolumn{9}{c}{\textbf{Panel (b)} $\Delta \%$ in EU Share of Domestic Purchases} \\
\textit{Total} 
  & \textbf{0.04} & & 0.04 
  & & \textbf{0.10} & & 0.10 \\
\quad \dots \textit{of clean intermediate goods} 
  & \textbf{0.08} & & 0.08 
  & & \textbf{0.18} & & 0.18 \\
\quad \dots \textit{of dirty intermediate goods} 
    & \textbf{0.03} & & 0.03 
  & & \textbf{0.07} & & 0.07 \\
\midrule
\multicolumn{9}{c}{\textbf{Panel (c)} $\Delta \%$ in EU Domar Weights} \\
\textit{Total} 
  & \textbf{0.03} & \textbf{0.19} & 0.03 & 0.19 
  & \textbf{0.14} & \textbf{0.40} & 0.14 & 0.39 \\
\quad \dots \textit{of clean intermediate goods} 
  & \textbf{0.11} & \textbf{0.09} & 0.10 & 0.09 & \textbf{0.29} & \textbf{0.29} & 0.28 & 0.28 \\
\quad \dots \textit{of dirty intermediate goods} 
  & \textbf{-0.03} & \textbf{0.22} & -0.02 & 0.22
  & \textbf{0} & \textbf{0.44} & 0 & 0.43\\
\bottomrule
\end{tabular}}
\caption{
\justifying
\scriptsize
Panel (a) presents the average percentage change across countries of the share that the EU purchases of intermediate inputs sourced from non-EU countries in EU total intermediate input purchases (\% of baseline year 2024). The standard deviations indicate heterogeneity of country-industry observations. Panel (b) shows the average percentage change of the share of intermediate inputs sourced domestically. Finally, Panel (c) reports the average percentage change in EU sectoral sales, with the corresponding standard deviations. Each value is considered under different scenarios: a full and a reduced version of CBAM can be combined with endogenous (Endog.) or exogenous (Exog.) prices. The original design of CBAM is the full version with endogenous prices.}
\label{tableIMPEXP}
\end{table}

Please note how reported effects are not that different across the endogenous and exogenous CBAM scenarios; that is, the additional adjustment in the endogenous scenario leads to a mere 5\% increase in the effect (from $0.57\%$ to $0.60\%$) under reduced CBAM, and a 3\% increase (from $1.23\%$ to $1.27\%$) under full CBAM. Based on this evidence, one might think that the effort to design a policy that is endogenous to carbon pricing is futile. Effects will be similar across all dimensions in the following paragraphs, when we show results on emissions and gains from trade. Yet, a caveat is needed. Our simulations are based on what happens in the current carbon market, in the absence of that strategic element that is crucial in the CBAM policy design. In fact, we have not yet observed any changes in technology by foreign producers in response to the actual CBAM, and there has been no establishment of additional national carbon markets by foreign partners. Both responses are foreseen in the original policy design. In this case, we believe our results for endogenous scenarios remain valid, and they can be considered the lower bound of what would happen in the absence of strategic responses. 

Let us describe how the feedback loop between the CBAM and equilibrium carbon prices operates in the current institutional environment. First, CBAM increases the relative price of foreign carbon-intensive goods; thus, demand shifts towards cleaner and domestic substitutes. This reallocation reduces production abroad and increases it at home, resulting in a decrease in foreign carbon prices due to lower demand for carbon-intensive inputs. Since the CBAM responds negatively to foreign carbon prices and positively to domestic ones, the differential widens, increasing the value of the CBAM. This general-equilibrium feedback magnifies the substitution away from foreign inputs. Hence, the albeit small magnification in panel (a), when moving from exogenous toward endogenous CBAM.

At a more granular level, while the CBAM leads to a contraction in the shares of foreign purchases of carbon-intensive sectoral goods directly targeted by the policy (with a 1.06\% decrease under the reduced CBAM, reaching 2.14\% under the full CBAM), there is a reallocation of the input demand towards cleaner alternatives driven by substitution effects, with an increase in the average share of foreign purchases of cleaner goods by 0.57\% under the reduced CBAM scenario, expanding to 1.91\% under the full CBAM. Please note how the magnitude of these substitution effects clearly depends on the elasticity of substitution across intermediate inputs, captured with $\theta$. As shown in Appendix Table \ref{tableIMPEXP_elasticities}, a higher elasticity amplifies the reallocation from dirty to cleaner inputs.

Panel (b) offers a complementary perspective to Panel (a). It presents the effect on the share of intermediate inputs sourced within the EU. The additional cost introduced by the CBAM leads to a partial substitution away from more expensive foreign intermediate inputs in favour of domestically produced goods, more so for clean than for dirty inputs.  Under the reduced CBAM, domestic purchases of clean goods rise by 0.08\%, while those of polluting goods increase by 0.03\%. These effects widen under the full CBAM, with domestic purchases of clean goods rising to 0.18\% and those of polluting goods to 0.07\%. 

Finally, panel (c) in Table \ref{tableIMPEXP} shows the average change in EU Domar weights (defined in Section \ref{sec:model_io_def}). This is necessary to understand how the size of the EU economy changes. Despite the small albeit positive overall change, disaggregated results reveal a difference between clean and dirty goods. The clean sectors slightly expand, and the dirty sectors slightly shrink. The average Domar weight of clean sectors exhibits an increase of 0.11\% under the reduced CBAM (0.29\% under the full CBAM), whereas the corresponding share of dirty goods declines marginally under the reduced CBAM and remains unchanged under the full CBAM. This asymmetric effect stems from the different reliance of clean and dirty producers on carbon-intensive inputs: while the CBAM raises unit production costs for all producers, clean producers are less exposed to this shock due to their lower input carbon intensity. As a result, their relative competitiveness increases, while European carbon-intensive sectors suffer a loss of sales. In fact, importers in other countries are more likely to reallocate their demand away from the European Union and towards third-country producers. Overall, the total EU Domar weights slightly increase, thus indicating a modest growth of the EU contribution to global production networks.

\begin{figure}[htb!]
    \centering
    \begin{subfigure}[t]{0.9\textwidth}
        \centering
        \includegraphics[width=\linewidth]{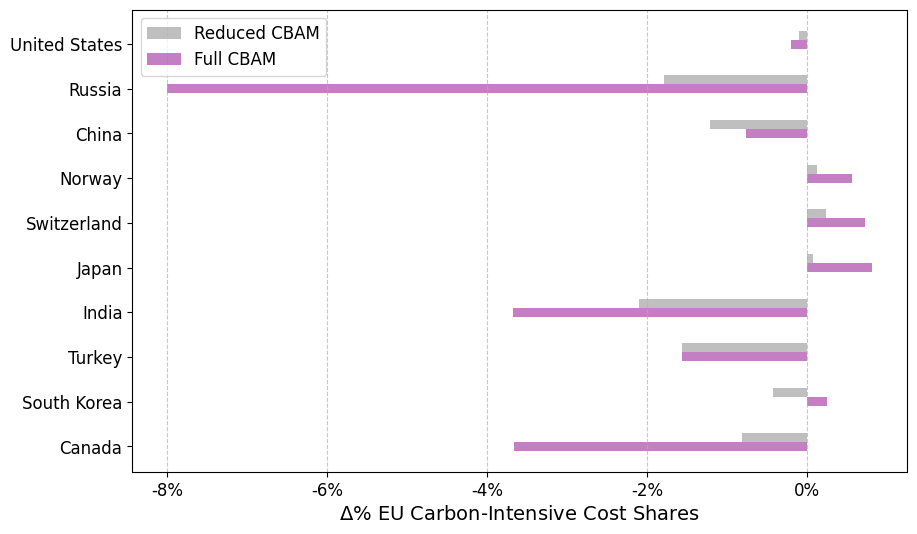}
        \caption{10 biggest exporters of polluting goods}
        \label{figCOSTSHARESCBAM1}
    \end{subfigure}
    \vspace{10mm}
    \begin{subfigure}[t]{0.9\textwidth}
        \centering
        \includegraphics[width=\linewidth]{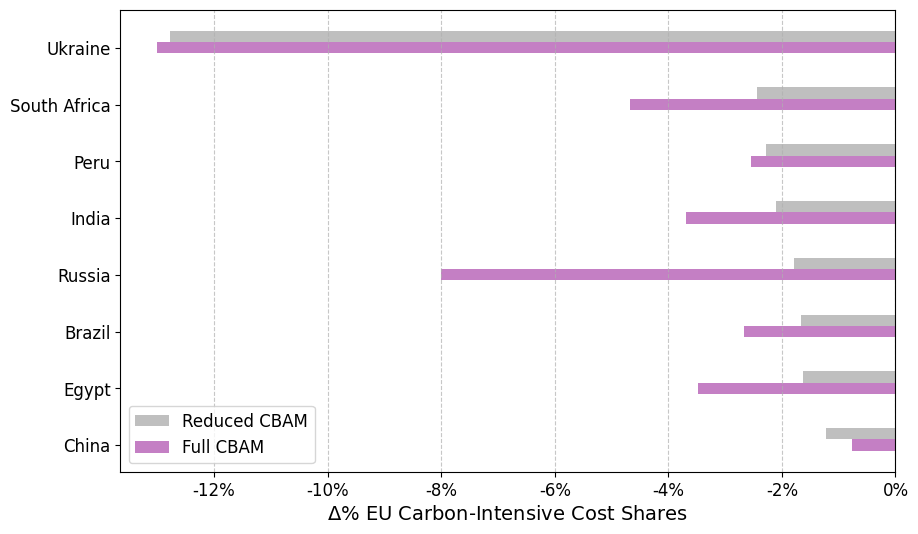}
        \caption{10 most affected countries}
        \label{figCOSTSHARESCBAM2}
    \end{subfigure}
    \caption{\justifying Changes in European shares of foreign purchases of dirty goods for selected countries (\% of baseline year 2024).}
    \label{figCOSTSHARESCBAM}
    \vspace{5ex}
\end{figure}

Notably, the effect of CBAM is heterogeneous across trade partners. This is shown by the standard deviations of Table \ref{tableIMPEXP}. The magnitude of the border adjustment varies across countries and sectors due to three key factors: (i) the carbon intensity of the production process; (ii) the relative stringency of carbon pricing; (iii) the sectoral composition of exports to the EU. These heterogeneous effects are illustrated in Figure \ref{figCOSTSHARESCBAM}, which shows the change in the share of carbon-intensive foreign purchases for the ten largest exporting countries of polluting inputs to the EU (panel \ref{figCOSTSHARESCBAM1}) \footnote{See Figure \ref{share_EU_imports_appendix} for the baseline shares of EU Carbon Intensive Imports by country.}. Interestingly, the introduction of the CBAM leads to an increase in shares of purchases from some of these countries. 

The increase in the share from Switzerland and Norway is straightforward to interpret: since the CBAM does not apply to imports from countries that are part of the EU-ETS, although they are not members of the EU (as in the case of Switzerland), their exports are not subject to the additional carbon price wedge. Briefly, producers located in non-EU countries within the European Economic Area benefit from a relative cost advantage compared to exporters from countries targeted by the CBAM, which translates into increased export shares to the EU. The latter is a significant side effect of CBAM that has been largely overlooked so far.

The cases of Japan and South Korea are different. Both countries are major exporters of carbon-intensive goods to the EU, yet they exhibit distinct patterns in their trade relations with other partners. This is because Japan has a relatively low carbon intensity in the production of carbon-intensive goods (see Appendix Table \ref{tab:rhos_appendix}), while South Korea has a relatively high carbon price (see Appendix Figure \ref{figECR_appendix}). Both factors lead to low CBAM relative to these countries. On the other hand, when we consider changes in the shares of foreign purchases, Ukraine is the most affected country, followed by South Africa, India, Russia, and China\footnote{Please note how our results align with the Relative CBAM Exposure Index reported by the World Bank. All countries identified as highly exposed to the CBAM in the index experience a decline in their export share to the European Union.}. 

\begin{figure}[htb]
    \centering
    \includegraphics[width=\textwidth]{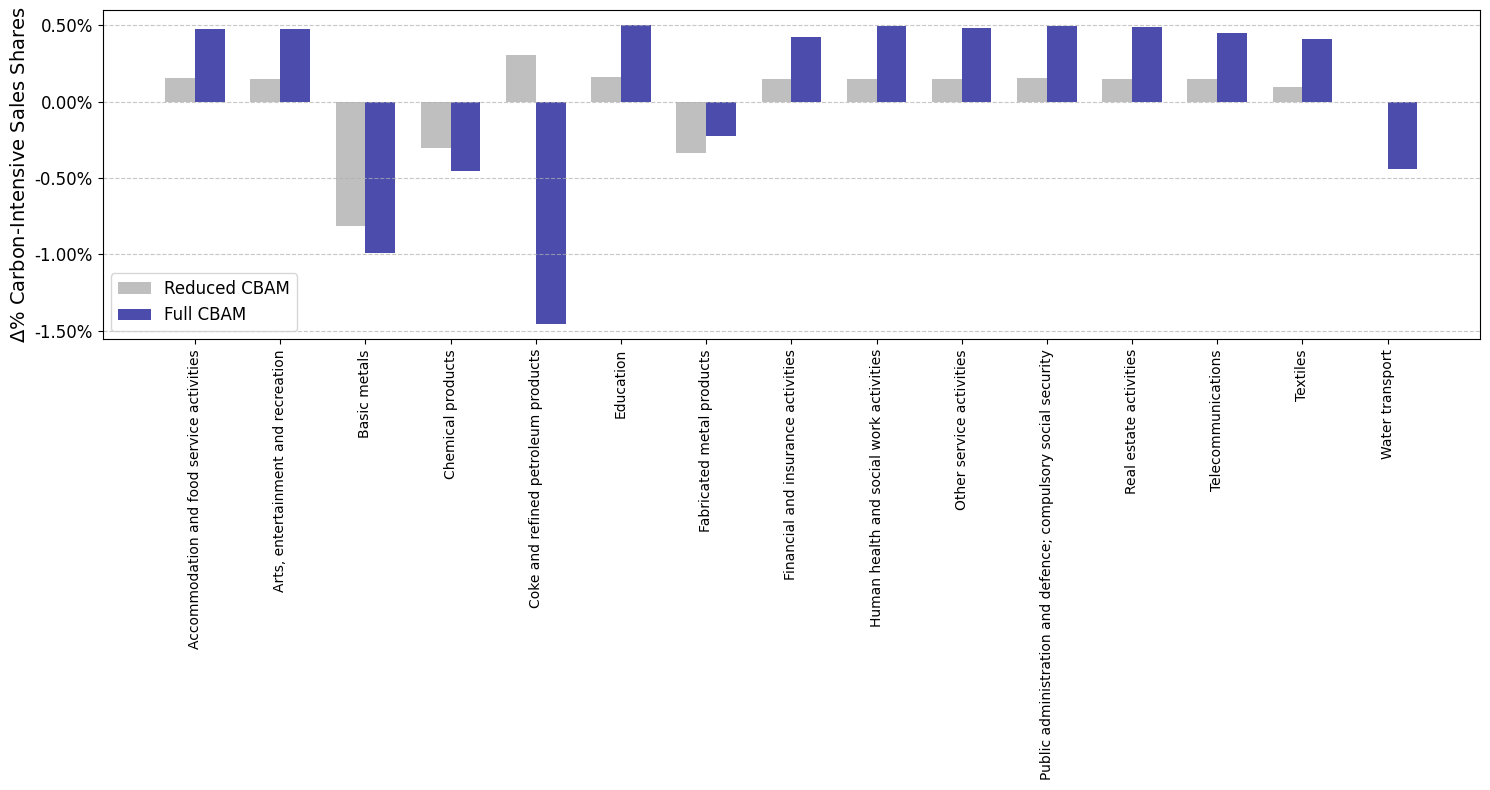}
    \caption{\justifying Changes in European Domar weights of selected sectors (\% of baseline year 2024).}
    \label{figSALESSHARESCBAM}
\end{figure}

Finally, Figure \ref{figSALESSHARESCBAM} illustrates the heterogeneous impact of CBAM on the EU Domar weights, highlighting the 15 most affected industries across both carbon-intensive and non-carbon-intensive categories. The most substantial declines occur in carbon-intensive sectors --- such as basic metals, chemical products, and fabricated metal products. In contrast, non-carbon-intensive sectors, which are not directly targeted by the reduced CBAM, experience marginal or modest gains. Such gains arise from substitution patterns along the supply chain, as captured by our model. As an illustration, consider \textit{Coke and refined petroleum products}. When only a subset of sectors is targeted under the reduced CBAM, \textit{Coke and refined petroleum products}, that is not subject to the wedge, act as a substitute for regulated inputs, experiencing an increase in its size. However, when the CBAM is extended to cover all sectors under the EU-ETS, including \textit{Coke and refined petroleum products}, the effect is reversed and the competitive advantage of the latter sector vanishes. These indirect effects become increasingly significant as the elasticity of substitution increases, as shown in sensitivity checks in Appendix Figure \ref{figSALESSHARESCBAM8}.

\subsection{Emissions.}



We start by adapting the definition of emissions from the OECD \parencite{yamano2020co2} to our theoretical framework. Let us consider the following equation, which indicates the emission embodied in European imports:

\begin{equation}\label{eq:EEI}
    \mathbf{EEI}(EU) \equiv \bm{\tilde{\rho}} (\mathbf{I} - \mathbf{\Omega})^{-1}\mathbf{X}_{EU} \mathbf{1}  = \bm{\tilde{\rho}} (\mathbf{I} -\mathbf{\Omega})^{-1} \gamma \mathbf{\tilde{\Omega}}_{EU} diag(\mathbf{\Lambda}) \mathbf{1} GNE
\end{equation}

Vector $\mathbf{EEI}(EU) \in \mathbb{R}^{NJ}$ has entries corresponding to the emissions embodied in European imports, as they are generated from the production of each good $j \in \mathcal{J}$ in every country $i \in \mathcal{N}$ embodied in European imports. In particular, the first equivalence in eq. \eqref{eq:EEI} corresponds to the OECD definition, while the subsequent equality restates it in terms of our theoretical framework. The term $\bm{\tilde{\rho}} = diag (\tilde{\rho}_1^1, ..., \tilde{\rho}_N^J) \in \mathcal{M}(NJ \times NJ)$ is a diagonal matrix with entries $\tilde{\rho}_i^j \equiv e_i^j /p_i^j q_i^j \ \forall i \in \mathcal{N}, j \in \mathcal{J}$ corresponding to the tons of emissions per dollar of output in a given sector-country pair. The matrix $\mathbf{\tilde{\Omega}}_{EU} \in \mathcal{M}(NJ \times NJ)$ contains input-output coefficients, with non-zero entries only for European destination sectors, excluding intra-EU trade. The vector $\mathbf{\Lambda} \in \mathbb{R}^{NJ}$ contains Domar weights, and $GNE$ is the world Gross National Expenditure. The term $\gamma \mathbf{\tilde{\Omega}}_{EU} diag(\mathbf{\Lambda}) \mathbf{1} GNE$ corresponds to the total value of European imports, expressed compactly as $\mathbf{X}_{EU} \mathbf{1} $, where $\mathbf{X}_{EU} \in \mathcal{M}(NJ \times NJ)$ is the matrix of trade flows between all country-sector pairs, with non-zero entries only for European destination sectors, once again excluding intra-EU trade. A more detailed derivation is left to the corresponding section in the Appendix \ref{appendix_proofs}. 

Briefly, The expression captures the total emissions embodied in European imports, including both emissions generated from the production of goods exported directly to the EU (emissions embodied in \textit{direct} imports) and emissions generated throughout the supply chain (emissions from \textit{indirect imports}), accounted for by the Leontief inverse $(\mathbf{I} - \mathbf{\Omega})^{-1}$. If we abstract from the upstream linkages by setting $\mathbf{\Omega} = \mathbf{0}$, only emissions embodied in \textit{direct} imports remain, reflecting emissions from goods produced abroad and directly exported to the EU.

\begin{table}
\centering
\scalebox{0.85}{
\begin{tabular}{@{}l@{\hskip 6pt}c@{}c@{\hskip 6pt}c@{}c@{}}
\toprule
\multirow{2}{*}{\textbf{Variable}} 
  & \multicolumn{2}{c}{\textbf{Reduced CBAM}} 
  & \multicolumn{2}{c}{\textbf{Full CBAM}} \\
  \cmidrule(l{6pt}r{0pt}){2-3} \cmidrule(l{6pt}r{0pt}){4-5}
& \textbf{\textit{Endog.}} & \textit{Exog.} & \textbf{\textit{Endog.}}& \textit{Exog.}  \\
\midrule
\multicolumn{5}{c}{\textbf{Panel (a) }$\Delta \% $ tons emissions embodied in direct imports} \\
\textit{Total} & \textbf{-4.73} & -4.70 & \textbf{-8.84} & -8.76 \\
\quad \dots \textit{of clean intermediate goods} & \textbf{0.87} & 0.85 & \textbf{2.54} & 2.48 \\
\quad \dots \textit{of dirty intermediate goods} & \textbf{-6.84} & -6.69 & \textbf{-13.12} & -12.84 \\
\midrule
\multicolumn{5}{c}{\textbf{Panel (b) }$\Delta \% $ tons emissions embodied in direct and indirect imports} \\
\textit{Total} & \textbf{-2.96} & -2.90 & \textbf{-5.19} & -5.08 \\
\quad \dots \textit{of clean intermediate goods} & \textbf{0.91} & 0.89 & \textbf{2.62} & 2.56 \\
\quad \dots \textit{of dirty intermediate goods} & \textbf{-3.94} & -3.86 & \textbf{-7.18} & -7.02 \\
\midrule
\multicolumn{5}{c}{\textbf{Panel (c) }$\Delta \% $ Emissions Leakage}\\
\textit{Total} &  \textbf{-0.07} & -0.07 & \textbf{-0.19} & -0.18 \\
\bottomrule
\end{tabular}}
\caption{\justifying Changes in emissions embodied in EU imports, and emissions leakage (\% of baseline year 2024). Each value is considered under two scenarios: one in which the CBAM is determined endogenously (Endog.) and one in which it is exogenous (Exog.)}.
\label{tableEEI}
\end{table}

Table \ref{tableEEI} presents the changes in emissions associated with European imports. Our results in panel (a) show that full CBAM leads to an 8.84\% reduction in emissions embodied in direct imports, reaching 13.12\% when only dirty intermediate goods are considered. When the indirect emissions are incorporated (panel (b)), the magnitude of reductions is attenuated but remains substantial: total emissions under the full CBAM reduce by 5.19\%. This attenuation is expected. In fact, while the CBAM directly reduces imports of carbon-intensive goods explicitly covered by the adjustment mechanism -- captured by the reduction in emissions embodied in direct imports -- it simultaneously induces substitution towards non-targeted inputs. This effect propagates upstream through the supply network, increasing the demand for upstream inputs, some of which are dirty, thereby raising indirect emissions. These upstream emissions are not accounted for in the design of CBAM. Briefly, failing to account for higher-order network effects results in a significant overestimation of CBAM efficacy, which is a novel result of our paper. As shown in Appendix Table \ref{tableEEI_elasticities}, a higher elasticity amplifies the effects.

Finally, panel (c) reports changes in emissions leakage. In our framework, the leakage coincides with the change in emissions of countries that have not yet adopted a national carbon market. Emissions leakage declines by 0.07\% under the reduced CBAM and by 0.19\% under the full CBAM. Thus, leakage still exists after CBAM, and it is only partially impacted by the border adjustment mechanism. See also the comments in Section \ref{sec:trade patterns}. Given the current institutional setting, we are unable to observe the strategic responses to the CBAM, which can help converge on policy targets.


Notably, we propose a decomposition of the total change in emissions embodied in European imports into two key components: 
\begin{enumerate}
    \item a \textit{technology} effect, which captures changes in emissions from production,
    \item a \textit{reallocation} effect, which captures changes in sourcing patterns across countries and sectors.
\end{enumerate}
A detailed derivation of the decomposition is provided in \ref{appendix_proofs}. For our current purposes, it is sufficient to note that the \textit{technology effect} corresponds to the changes in the vector $\bm{\tilde{\rho}}$ of emissions per dollar of output in equation (\ref{eq:EEI}), holding the structure of global production linkages constant. Unlike $\rho_i^j =t_ie_i^j/p_i^jq_i^j$, which is a fixed parameter of our model, $\tilde{\rho}_i^j =e_i^j/p_i^jq_i^j$ varies in response to changes in the input mix. Intuitively, this captures the reduction in the tons of emissions generated from production resulting from a shift toward cleaner inputs, given existing production chains. By contrast, the \textit{reallocation effect} stems from changes in the production network, captured by the Leontief inverse $(\mathbf{I} - \mathbf{\Omega})^{-1}$, and in the European imports, represented by $\mathbf{X}_{EU}$, holding production of emissions fixed. 

\begin{figure}[htb]
    \centering
    \includegraphics[width=0.8\linewidth]{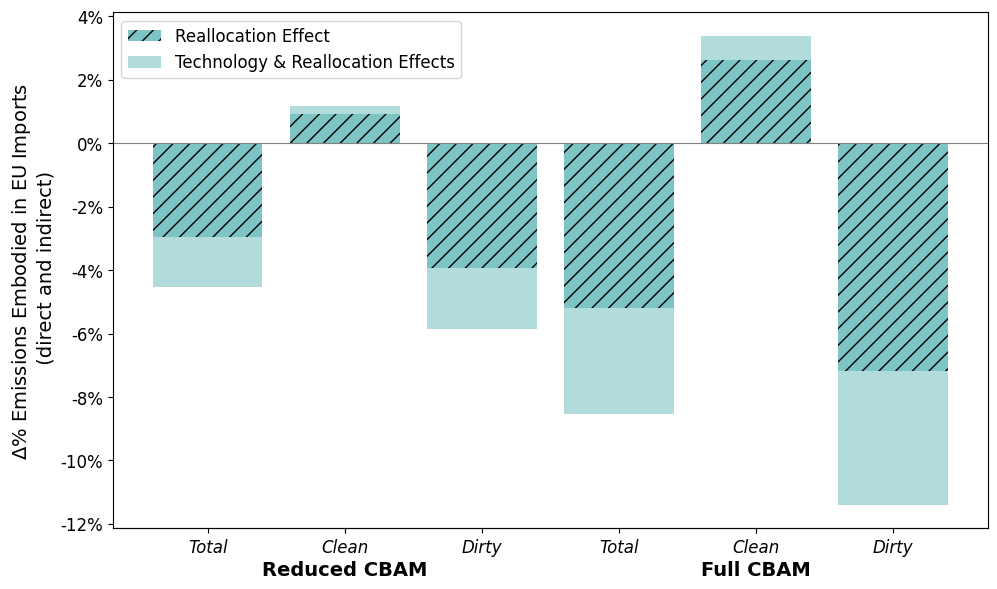}
    \caption{ \justifying Decomposition of the change in Emissions Embodied in European Imports.}
    \label{figDecompositionEEI}
\end{figure}

The corresponding Figure \ref{figDecompositionEEI} illustrates the results of the decomposition. In all cases -- whether total, clean, or dirty imports, and under both the reduced and full CBAM scenarios -- the reallocation effect accounts for more than 50\% of the total change in emissions. This evidence underscores the dominant role of trade reallocation mechanisms and highlights the importance of tracking changes that occur across the production network in order to understand the environmental impact of carbon border adjustments. In conclusion, we argue, it is important to adopt a supply-chain-wise perspective to ensure that full emissions' footprint is captured.


\subsection{Welfare gains.}

In the following paragraphs, we examine the effects of the CBAM on real GNE, as the latter proxy our welfare gains\footnote{Please note how welfare gains are driven by assumptions, among others, on fixed deficits and inelastic labor supply.}. Table \ref{tableGDPEMISSIONS} reports the percentage changes in EU (panel (a)) and extra-EU (panel (b)) real gross national expenditures following the introduction of the CBAM. Under the reduced configuration, EU GNE slightly increases by 0.006\%, while under the full CBAM scenario, the increase reaches 0.04\%. This modest improvement in the EU GNE is driven by the reallocation of demand away from foreign carbon-intensive goods and toward domestic production. Given that trade deficits are fixed in the model, this translates into an increase in the price of domestic goods relative to foreign ones (i.e., a terms of trade effect), which improves the EU trade balance and raises GNE. Please note how results are consistent with \textcite{florez2024eu}, who report a welfare gain of 0.016\%. Despite the increase in GNE, real wages decline by 0.02\% and 0.05\% under the reduced and full CBAM scenarios, respectively, as a consequence of a higher price index. 

Finally, panel (b) registers a moderate effect on GNE in extra-EU countries:  $-0.008\%$ under the reduced CBAM and $-0.02\%$ under the full CBAM scenario. Similar losses in real wages are also recorded in the same table. In this context, concerns about a disproportionate burden on developing or emerging economies seems exaggerated. However, please note that our theoretical setup abstracts from labor market or institutional frictions in the adjustment to the new equilibrium. In this case, it is also more evident that scenario with endogenous prices are not different from the ones with exogenous prices. As already discussed in Section \ref{sec:trade patterns}, our exercises do not consider possible strategic responses by partner countries, which are however an important element of the original CBAM design. The introduction of national carbon markets in extra-EU countries or the investment in less carbon-intensive technologies could help widen the gap between the exogenous and endogenous scenarios we report, and thus help converging to the target of lower emissions.

\begin{table}[hbt]
\centering
\scalebox{0.85}{
\begin{tabular}{@{}l@{\hskip 6pt}c@{}c@{\hskip 6pt}c@{}c@{}}
\toprule
\multirow{2}{*}{\textbf{Variable}} 
  & \multicolumn{2}{c}{\textbf{Reduced CBAM}} 
  & \multicolumn{2}{c}{\textbf{Full CBAM}} \\
  \cmidrule(l{6pt}r{0pt}){2-3} \cmidrule(l{6pt}r{0pt}){4-5}
& \textbf{\textit{Endog.}} & \textit{Exog.} & \textbf{\textit{Endog.}}& \textit{Exog.}  \\
\midrule
\multicolumn{5}{c}{\textbf{Panel (a) }$\Delta \% $ EU Gross National Expenditure} \\
\textit{Total} & \textbf{0.006} & 0.006 & \textbf{0.04} & 0.04 \\
\qquad  \dots \textit{Real Wages} & \textbf{-0.02} & -0.02 & \textbf{-0.05} & -0.05 \\
\midrule
\multicolumn{5}{c}{\textbf{Panel (b) }$\Delta \% $ extra-EU Gross National Expenditure} \\
\textit{Total} & \textbf{-0.008} & -0.008 & \textbf{-0.02} & -0.02 \\
\qquad \dots \textit{Real Wages} & \textbf{-0.009} & -0.008 & \textbf{-0.02} & -0.02 \\
\bottomrule
\end{tabular}}
\caption{\justifying Policy-induced changes in Gross National Expenditure (\% of baseline level) as a proxy for welfare. Each value is considered under two scenarios: one in which the CBAM is determined endogenously (Endog.) and one in which it is exogenous (Exog.).}
\label{tableGDPEMISSIONS}
\end{table}

\subsection{Separating trade and technology.}
\label{sec:synthetic_network}

Finally, we present two counterfactual exercises to shed light on how technology, on the one hand, and international trade, on the other hand, can influence the effects of CBAM. In the first, we aim to capture the effect of technology; therefore, we vary the output elasticity with respect to emissions. In the second part, we examine the role of the EU's economic integration within global supply chains. In both exercises, we focus on the key outcome of our analysis --- embodied emissions in EU imports. 

\begin{figure}
    \centering
    \includegraphics[width=0.8\linewidth]{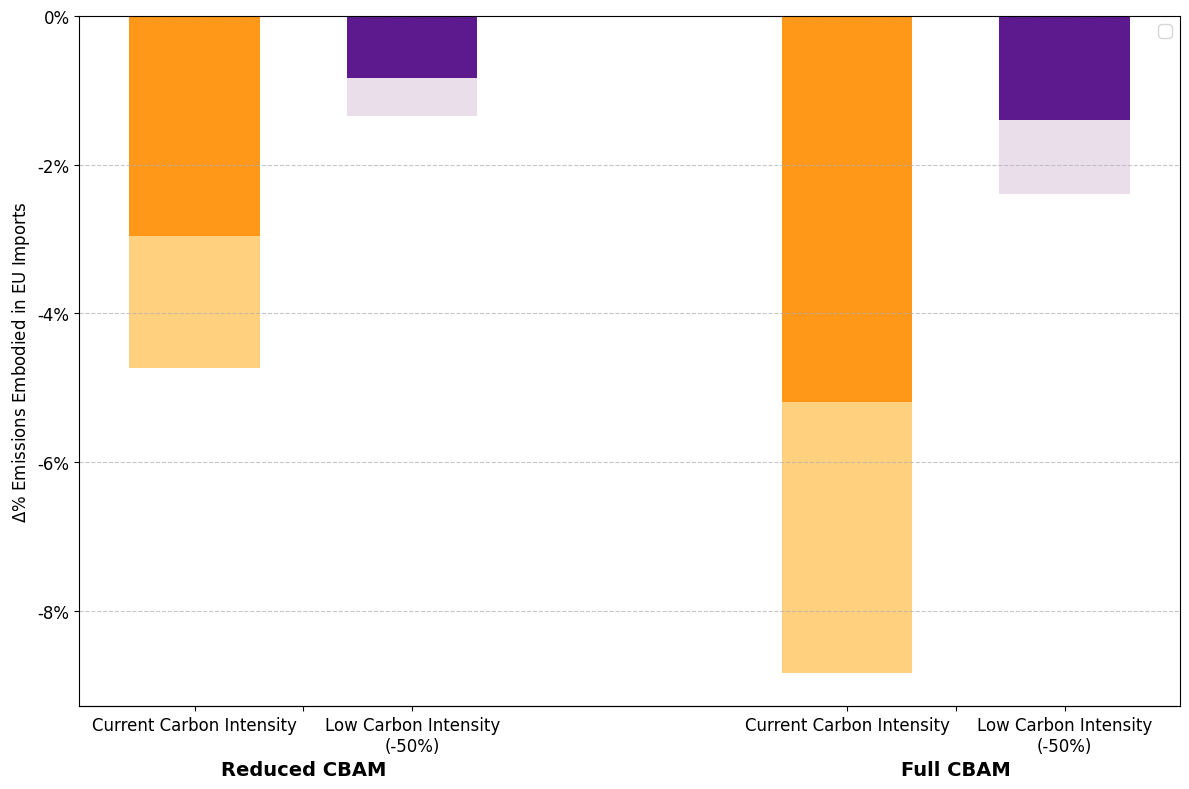}
    \caption{\justifying 
    \scriptsize 
    Changes in emissions embodied in EU imports under current and lower carbon intensity are shown with bars. Carbon intensities are lower when they are half of their reference value ($-50\%$). Percentages are computed with reference to the value of the corresponding embodied emissions (direct or total) under the scenario with current carbon intensities and no CBAM. Darker colors indicate the amount of emissions in direct imports, while lighter colors indicate the amount of emissions in indirect imports. }
    \label{fig:counterfactuals_rhos}
\end{figure}

Let us start with the effect of technology. We construct a counterfactual in which carbon intensities, captured by $\rho_i^j$, are just half ($-50\%$) of those observed at the baseline. This is a circumstance when the world's technology becomes cleaner. Figure \ref{fig:counterfactuals_rhos} shows the reduction in emissions embodied in EU imports—both direct and total—under the two scenarios. Results are presented as percentage changes relative to the baseline. As expected, the impact of the CBAM on reducing embodied emissions weakens in a world where global production is already cleaner, i.e., less carbon-intensive, since the associated CBAM charges are lower. Interestingly, the reduction is more than proportional, underscoring the disproportionate impact of production technologies on the policy’s effectiveness.

\begin{figure}[htb!]
    \centering
    \begin{subfigure}[t]{0.8\textwidth}   
        \centering
        \includegraphics[width=\linewidth]{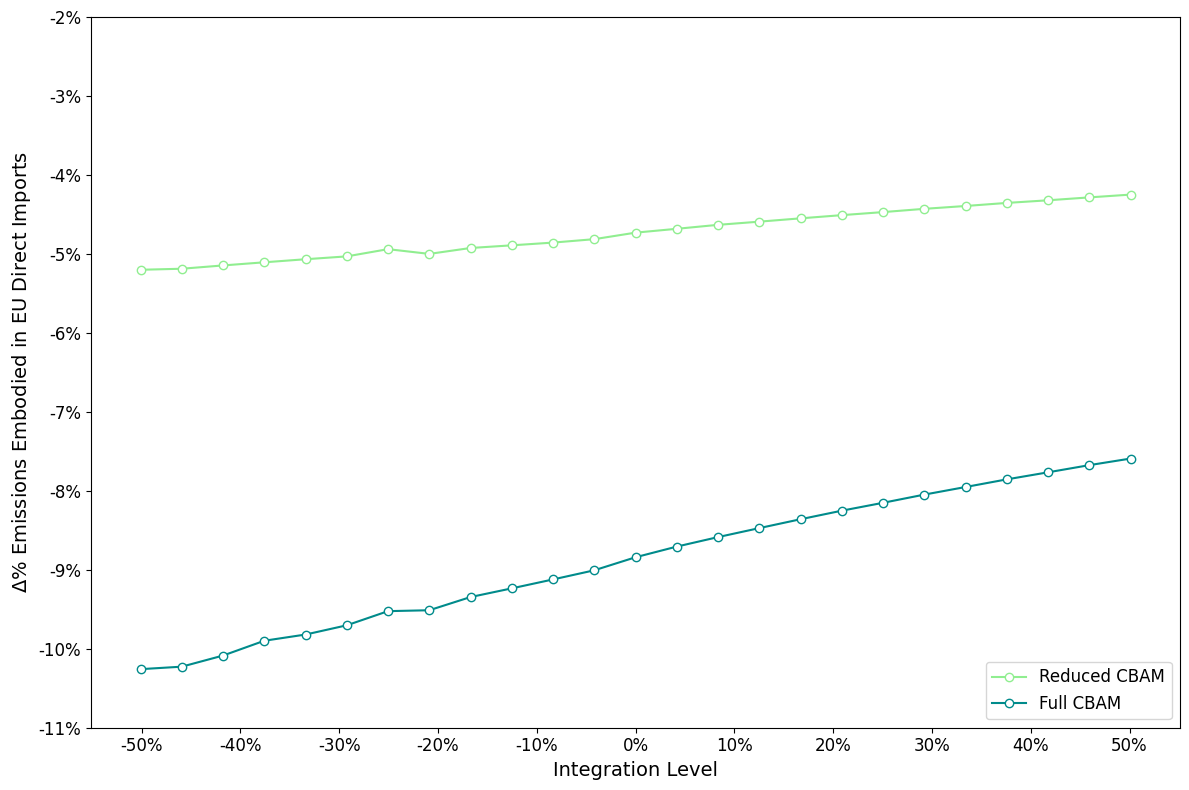}
        \caption{Emissions embodied in direct imports}
        \label{fig:EEI1_integration}
    \end{subfigure}
    \hspace{10mm}
    \begin{subfigure}[t]{0.8\textwidth}
        \includegraphics[width=\linewidth]{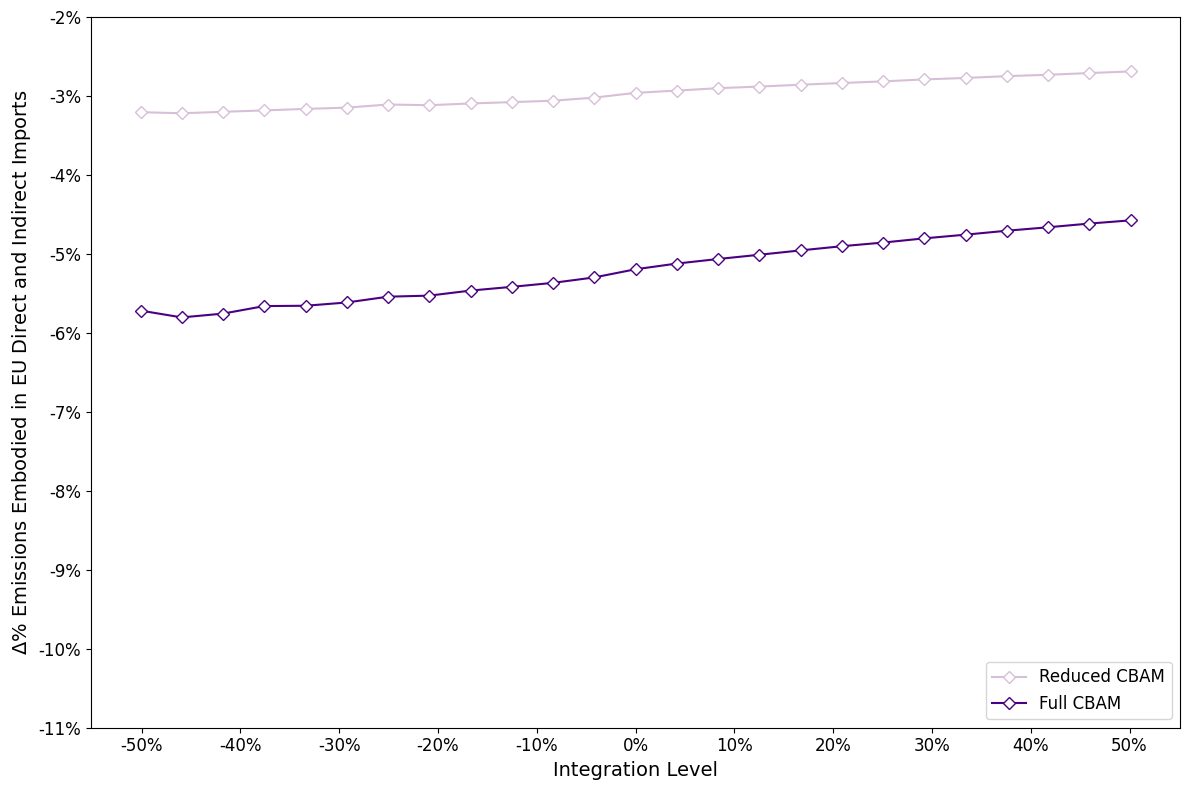}
        \caption{Emissions embodied in direct and indirect imports}
        \label{fig:EEI2_integration}
    \end{subfigure}
    \caption{\justifying Changes in emissions embodied in EU Imports under varying levels of trade integration Percentages are computed with reference to the value of the corresponding embodied emissions (direct or total) under the scenario with the current level of trade integration (0\% change) and no CBAM.}
    \label{fig:counterfactuals_integration}
\end{figure}

Let us now assess the role of international trade, which implies the construction of a counterfactual where we systematically alter the reliance on imported inputs in a range $[-50\%, +50\%]$. Specifically, for each EU importing sector, we vary the ratio of imported to domestic inputs, where a higher ratio reflects deeper integration of EU production into global supply chains. We then plot the impact of the CBAM on embodied emissions in Figure \ref{fig:counterfactuals_integration} as a function of the foreign inputs ratio, while holding all else constant. CBAM induces two types of substitution: towards domestic inputs and towards cleaner non-targeted foreign inputs. We find that when the EU economy is less integrated -- that is, when sectors rely more on domestic rather than imported inputs -- the CBAM leads to a stronger shift toward domestic sourcing. This reallocation results in a larger reduction in emissions embodied in imports, simply because a greater share of inputs originates within the EU and thus lies outside the CBAM scope. As integration increases, the dependence on foreign inputs rises, diminishing the potential for substituting with domestic inputs, and consequently weakening the emissions-reducing effect of the policy. Nevertheless, the CBAM continues to influence sourcing decisions, prompting a shift toward cleaner goods, which partly offsets the reduced scope for domestic substitution. Although emissions embodied in imports still fall relative to the baseline, the overall reduction is smaller, highlighting the dominant role of domestic substitution in driving the policy's impact.

We conclude that our counterfactuals demonstrate how cleaner global production considerably reduces the scope for CBAM, underscoring the importance of technological progress. On the other hand, a higher level of supply chain integration weakens policy effectiveness, reducing the scope for substitution toward domestic inputs. Eventually, international trade has an impact on CBAM, albeit to a lesser extent than technological progress.  

\section{Conclusion}\label{sec:conclusion}
This paper provides a comprehensive assessment of the economic and environmental implications of the EU’s Carbon Border Adjustment Mechanism, utilizing a multi-country, multi-sector general equilibrium model with global input-output linkages. By endogenizing both the CBAM and carbon prices, we capture the feedback mechanisms that arise between national climate policies and international trade patterns. In line with the original policy design, we present a $ 2\times 2$ alternative scenario to illustrate its implementation. A full and a reduced CBAM scenario are combined with the possibility of endogenizing or not CBAM prices, respectively. Results challenge the notion of a trade-off between eliminating carbon leakage, on the one hand, and gains from welfare and trade, on the other. We find that CBAM crucially induces a reallocation of purchases towards domestic and foreign clean inputs, and significantly reduces emissions embodied in imports. Moreover, a positive albeit small increase is detected for EU Gross National Expenditures (GNE), which is primarily due to a terms-of-trade effect. Finally, counterfactual exercises demonstrate the primary roles of technological progress (i.e., cleaner production) and trade (i.e., increased foreign inputs) in emission reduction, where the first dominates.

Crucially, our findings highlight the importance of supply-chain-wise measures to capture the full carbon footprint in production. Moreover, although it is beyond the scope of this paper to evaluate the feasibility of a CBAM design, our efforts in collecting data have made clear the importance of having an extensive information system capable of collecting and regularly updating data on the emissions generated worldwide by various products. The transitional phase is approaching its end, and it is unclear whether the policy can be in force, as planned, by 2026.
More generally, our paper contributes to the study of solutions to international coordination problems when economies are independent and full enforcement cannot be achieved in the face of imminent global challenges, including climate change. We can think of other different international coordination problems (international taxation, trade policy, etc.) that, like environmental regulations, necessitate policies that account for the complex interdependence of global production networks.

\newpage
\printbibliography

\appendix
\renewcommand{\thesection}{\Alph{section}}
\renewcommand{\thesubsection}{\Alph{section}.\arabic{subsection}}
\newpage
\section{Appendix}

\subsection{Data}\label{appendix_data}

\beginsupplement

The following section describes the data used in the quantitative analysis. The list of considered countries includes: Argentina, Australia, Bangladesh, Brazil, Canada, Chile, China, Colombia, Costa Rica, Cote d'Ivoire, Egypt, European Union, Iceland, India, Indonesia, Israel, Japan, Malaysia, Morocco, Mexico, New Zealand, Nigeria, Norway, Peru, Philippines, Russia, South Africa, South Korea, Switzerland, Turkey, Ukraine, United States and an aggregate Rest of the World\footnote{\ Due to the unavailability of data on emissions and emissions pricing, we move the following countries to the "Rest of the World" category in our analysis: Brunei, Cameroon, Hong Kong, Jordan, Kazakhstan, Cambodia, Laos, Myanmar, Pakistan, Saudi Arabia, Senegal, Singapore, Thailand, Tunisia, Taiwan, and Vietnam.}. The list of sectors is reported in Table \ref{tableA1}, where we aggregate the two sectors of "Other service activities" and "Activities of households as employers; undifferentiated goods- and services-producing activities of households for own use" under the 44th sector in order to eliminate zero domestic production shares from the Input-Output table.

\begin{itemize}
    \item\textbf{Input-Output Tables}
    To calibrate the model on the year 2018, we use the 2021 release of the OECD Inter-Country Input-Output tables \parencite{oecd2021}, collecting data on expenditures in intermediate inputs $X_{ni}^{kj}$ and final goods $F_{ni}^j$, gross output $GO_i^j$ and value-added $VA_i^j$. Concerning final consumption, we ignore Gross Fixed Capital Formation and Changes in Inventories and Valuables. Values are reported in millions of U.S. dollars at current prices and the flows of goods and services within and across countries are directly mapped to the cost shares in intermediate goods $\tilde{\omega}_{ni}^{kj} = X_{ni}^{kj}/\sum_{n \in \mathcal{N}} \sum_{j \in \mathcal{J}} X_{ni}^{kj}$ and the shares of final consumption $\alpha_i^j = \sum_{n \in \mathcal{N}} F_{ni}^j / \sum_{n \in \mathcal{N}} \sum_{j \in \mathcal{J}} F_{ni}^j$. Finally, the shares of value added are computed as $\beta_i^j = VA_i^j/GO_i^j$. We have a unique country-sector pair (Chile - Motor vehicles, trailers and semi-trailers) with a gross output equal to 0, which can be either due to missing observations or zero production in that sector and country. Since these cases generate computational issues, we input a value of 1 to this observation. 
    
    \item \textbf{Tariffs} Bilateral tariffs for the year 2018 are collected from the United Nations Statistical Division-Trade Analysis and Information System (UNCTAD-TRAINS). We consider effective applied rates at 6-digit of the Harmonised System and build the weighted average tariff corresponding to our sectoral classification, using the associated import values. For the cases in which a single HS6 code corresponds to more than one ISIC 2-digit code, we impute the same share of import values to all the ISIC categories. Finally, when tariff data for the year 2018 were not available, we substitute this value with the closest one available, searching for the three previous years.   

    \item \textbf{Emissions and carbon prices} Emissions from production are taken from the OECD Environmental Statistics database. We consider Scope 1 Emissions for each country-sector pair $E_i^j$, including all direct GHG emissions measured in million tonnes of $\text{CO}_2$ equivalent. For the European Union, Iceland and Norway we employ a distinct approach. Emissions data are taken from the European Union Transaction Log (EUTL), recording all transactions carried out under the EU-ETS, with information on each entity covered by the system. The EUTL provides granula, entity-level information on verified emissions and the volume of freely allocated allowances on an annual basis. We aggregate these data at the selected sectoral level and adjust the reported emissions by deducting those covered by freely allocated allowances. To calculate the emissions elasticity, $\rho_i^j$, these values are multiplied by the corresponding country's Effective Carbon Rate (ECR), which is then converted into USD using the OECD exchange rate for the year 2018. Effective Carbon Rates represent the total price $t_i$ of one ton of $\text{CO}_2$ emissions as the sum of carbon taxes, specific taxes on energy use, and the price of tradable emission permits. Emission elasticities are then computed as $\rho_i^j = t_iE_i^j/GO_i^j$.  
\end{itemize}

\renewcommand{\arraystretch}{1}
\begin{table}[p]
\footnotesize
    \centering
    \scalebox{0.9}{\begin{tabular}{p{8.5cm} l l c c}
    \midrule
    Industry Description & OECD Code & ISIC Rev.4 & ETS & CBAM \\
    \midrule
    Agriculture, hunting, forestry & $\text{A01\_02}$ &  01, 02 & 0 & 0 \\
    Fishing and aquaculture & A03 & 03 & 0 & 0\\
    Mining and quarrying, energy producing products & $\text{B05\_06}$ & 05, 06 & 1 & 0\\ 
    Mining and quarrying, non-energy producing products & $\text{B07\_08}$ & 07, 08 & 1 & 1 \\
    Mining support service activities & B09 & 09 & 1 & 0 \\  
    Food products, beverages and tobacco &  $\text{C10\_12}$ & 10 to 12 & 1 & 0\\
    Textiles, textile products, leather and footwear & $\text{C13\_15}$ & 13 to 15 & 0 & 0\\
    Wood and products of wood and cork & C16 & 16 & 1 & 0\\
    Paper products and printing & $\text{C17\_18}$ & 17, 18 & 1 & 0\\
    Coke and refined petroleum products &  C19 & 19 & 1 & 0\\
    Chemical and chemical products & C20 & 20 & 1 & 1\\
    Pharmaceuticals, medicinal chemical and botanical products & C21 & 21 & 1 & 0\\
    Rubber and plastics products & C22 & 22 & 1 & 0\\
    Other non-metallic mineral products & C23 &  23 & 1 & 1 \\
    Basic metals & C24 & 24 & 1 & 1\\
    Fabricated metal products & C25 & 25 & 1 &1\\
    Computer, electronic and optical equipment & C26 & 26 & 1 & 0\\
    Electrical equipment & C27 & 27 & 1 & 0 \\
    Machinery and equipment, nec & C28 & 28 & 0 & 0 \\
    Motor vehicles, trailers and semi-trailers & C29 & 29 & 1 & 0 \\
    Other transport equipment & C30 & 30 & 1 & 0\\
    Manufacturing nec; repair and installation of machinery and equipment & $\text{C31\_33}$ & 31 to 33 & 1 & 0 \\
    Electricity, gas, steam and air conditioning supply & D & 35 & 1 & 1\\
    Water supply; sewerage, waste management and remediation activities & E & 36 to 39 & 0 & 0 \\
    Construction & F & 41 to 43 & 1 & 0 \\
    Wholesale and retail trade; repair of motor vehicles & G & 45 to 47 & 1& 0\\
    Land transport and transport via pipelines & H49 & 49 & 0 & 0 \\
    Water transport & H50 & 50 & 0 & 0 \\
    Air transport & H51 & 51 & 0 & 0 \\
    Warehousing and support activities for transportation & H52 & 52 & 1 & 0 \\
    Postal and courier activities & H53 & 53 & 0 & 0 \\
    Accommodation and food service activities & I & 55, 56 & 0 & 0\\
    Publishing, audiovisual and broadcasting activities & $\text{J58\_60}$ & 58 to 60 & 0 & 0 \\
    Telecommunications & J61 & 61 \\
    IT and other information services & $\text{J62\_63}$ & 62, 63 & 0 & 0\\
    Financial and insurance activities & 64 to 66 & 1 & 0 \\
    Real estate activities & L & 68 & 1 & 0\\
    Professional, scientific and technical activities & M & 69 to 75 & 1 & 0\\
    Administrative and support services & N & 77 to 82 & 1 & 0 \\
    Public administration and defence; compulsory social security & O & 84 & 0 & 0\\
    Education & P & 85 \\
    Human health and social work activities & Q & 86 to 88 & 0 & 0 \\
    Arts, entertainment and recreation & R  & 90 to 93 & 0 & 0\\
    Other service activities + Activities of households as employers; undifferentiated goods- and services-producing activities of households for own use & S, T & 94 to 98 & 0 & 0\\ 
    \midrule
    \end{tabular}}
    \caption{The table lists the sectoral classification in our analysis, distinguishing whether a sector is covered by the ETS (\textit{ETS}=1), CBAM(\textit{CBAM}=1) or both.}
    \label{tableA1}
\end{table}

\pagebreak 

\subsection{The model}\label{appendix_proofs}

\paragraph{Marginal cost}

\begin{lemma}
    \textit{The marginal cost of a firm is given by:}
    
    \begin{equation}\label{eq:8}
        mc_i^j =  \frac{1}{(A_i^j)^{1-\rho_i^j}} \left[ w_i^{\beta_i^j} (P_i^j)^{1-\beta_i^j}\right]^{1- \rho_i^j} [t_i (1-\epsilon_i^j)]^{\rho_i^j}
    \end{equation}
    
\end{lemma}
\vspace{3mm}

 \textit{Proof.} Each producer solves the following cost minimisation problem:

$$\min_{l_i^j, e_i^j, z_{1i}^{1j}, ..., z_{Ni}^{Jj} } TC_i^j= w_il_i^j + \sum_{n \in \mathcal{N}} \sum_{k \in \mathcal{J}} p_{ni}^{kj} z_{ni}^{kj} + t_i (1-\epsilon_i^j) e_i^j $$
$$\qquad \text{ s.t. }  q_i^j \le \Upsilon_i^j \left[ A_i^j (l_i^j)^{\beta_i^j} \left( \sum_{n \in \mathcal{N}} \sum_{k \in \mathcal{J}} (\iota_{ni}^{kj})^{\frac{1}{\theta}} (z_{ni}^{kj})^{\frac{\theta-1}{\theta}}\right)^{\frac{\theta}{\theta-1}(1-\beta_i^j)} \right]^{1-\rho_i^j} [e_i^j]^{\rho_i^j}$$

 The Lagrangian is equal to:

$$\mathcal{L}(l_i^j, e_i^j, z_{1i}^{1j}, ..., z_{Ni}^{Jj}, \lambda) =  w_il_i^j + \sum_{n \in \mathcal{N}} \sum_{k \in \mathcal{J}} p_{ni}^{kj} z_{ni}^{kj} + t_i(1-\epsilon_i^j) e_i^j  + $$
$$ + \lambda \left\{  q_i^j - \Upsilon_i^j \left[ A_i^j (l_i^j)^{\beta_i^j} \left( \sum_{n \in \mathcal{N}} \sum_{k \in \mathcal{J}} (\iota_{ni}^{kj})^{\frac{1}{\theta}} (z_{ni}^{kj})^{\frac{\theta-1}{\theta}}\right)^{\frac{\theta}{\theta-1}(1-\beta_i^j)} \right]^{1-\rho_i^j} [e_i^j]^{\rho_i^j} \right\} $$
\vspace{3mm}

 and the FONCs, which are also sufficient given the assumed Cobb-Douglas production function, are:

$$
\begin{cases}
    \frac{\partial \mathcal{L}}{\partial l_i^j} = w_i - \lambda (1- \rho_i^j) \beta_i^j \frac{q_i^j}{l_i^j} = 0 \\
    \frac{\partial \mathcal{L}}{\partial e_i^j} = t_i(1-\epsilon_i^j) - \lambda \rho_i^j \frac{q_i^j}{e_i^j} = 0  \\
    \forall n, k: \quad \frac{\partial\mathcal{L}}{\partial z_{ni}^{kj}} = \frac{\partial \mathcal{L}}{\partial M_i^j} \frac{\partial M_i^j}{\partial z_{ni}^{kj}} =  p_{ni}^{kj} - \lambda \left[(1-\beta_i^j)(1-\rho_i^j) \frac{q_i^j}{M_i^j} \right] \left[ (\iota_{ni}^{kj})^{\frac{1}{\theta}} (z_{ni}^{kj})^{\frac{\theta-1}{\theta} - 1} (M_i^j)^{\frac{1}{\theta}} \right] = 0 \\
    \frac{\partial \mathcal{L}}{\partial \lambda} =  q_i^j - \Upsilon_i^j \left[ A_i^j (l_i^j)^{\beta_i^j} (M_i^j)^{1-\beta_i^j} \right]^{1-\rho_i^j} [e_i^j]^{\rho_i^j} = 0
\end{cases}
$$
\vspace{3mm}

 By taking the ratio of any two intermediate inputs and using the CES aggregate for materials $M_i^j$, the optimal amount of each intermediate input $z_{ni}^{kj}$ is given by:

\begin{equation}\label{eqINTINPUTS}
    z_{ni}^{kj} = \iota_{ni}^{kj} \left( \frac{p_{ni}^{kj}}{P_i^j} \right)^{-\theta} M_i^j
\end{equation}

  with $ P_i^j \equiv  \left( \sum_{n\in \mathcal{N}} \sum_{k \in \mathcal{J}} \iota_{ni}^{kj}(p_{ni}^{kj})^{1-\theta} \right)^{ \frac{1}{1-\theta}} $ being the price index of the intermediate goods. \\
Using (16) and the FOC for the intermediate inputs:

$$M_i^j = \lambda (1-\beta_i^j)(1-\rho_i^j) q_i^j (P_i^j)^{-1}$$

 and using the expression for $\lambda$ from the FOC for the labour $l_i^j$:

$$M_i^j = \frac{1-\beta_i^j}{\beta_i^j} \frac{w_i}{P_i^j} l_i^j$$

 Moreover, from the FOCs for labor and emissions we get:

$$e_i^j = \frac{\rho_i^j}{\beta_i^j(1-\rho_i^j)} \frac{w_i}{t_i(1-\epsilon_i^j)} l_i^j$$

 Plugging the above equations for $M_i^j$ and $l_i^j$ into the production function and solving for  $l_i^j$, $M_i^j$ and $e_i^j$, the conditional input demand functions are given by: 

$$
    l_i^j = \frac{1}{\Upsilon_i^j} \frac{q_i^j}{(A_i^j)^{1-\rho_i^j}} \left( \frac{\beta_i^j}{1-\beta_i^j} \frac{P_i^j}{w_i} \right)^{(1-\beta_i^j)(1-\rho_i^j)} \left( \frac{\beta_i^j(1-\rho_i^j)}{\rho_i^j} \frac{t_i(1-\epsilon_i^j)}{w_i}\right)^{\rho_i^j} =
$$    

\begin{equation}\label{eq:10}
    = \beta_i^j(1-\rho_i^j) \frac{q_i^j}{(A_i^j)^{1-\rho_i^j}} \left[w_i^{\beta_i^j(1-\rho_i^j)-1}  \left( P_i^j\right)^{(1-\beta_i^j)(1-\rho_i^j)} [t_i(1-\epsilon_i^j)]^{\rho_i^j} \right]
\end{equation}

$$
    M_i^j = \frac{q_i^j}{\Upsilon_i^j(A_i^j)^{1-\rho_i^j}} \left( \frac{\beta_i^j}{1-\beta_i^j} \frac{P_i^j}{w_i} \right)^{(1-\beta_i^j)(1-\rho_i^j) -1} \left( \frac{\beta_i^j(1-\rho_i^j)}{\rho_i^j} \frac{t_i(1-\epsilon_i^j)}{w_i}\right)^{\rho_i^j}
$$
\begin{equation}\label{eq:11}
    = (1-\beta_i^j) (1-\rho_i^j) \frac{q_i^j}{(A_i^j)^{1-\rho_i^j}} \left[ w_i^{\beta_i^j(1-\rho_i^j)} \left( P_i^j\right)^{(1-\beta_i^j)(1-\rho_i^j) -1} [t_i(1-\epsilon_i^j)]^{\rho_i^j} \right]
\end{equation}

$$
    e_i^j = \frac{q_i^j}{\Upsilon_i^j(A_i^j)^{1-\rho_i^j}} \left( \frac{\beta_i^j}{1-\beta_i^j} \frac{P_i^j}{w_i} \right)^{(1-\beta_i^j)(1-\rho_i^j)} \left( \frac{\beta_i^j(1-\rho_i^j)}{\rho_i^j} \frac{t_i(1-\epsilon_i^j)}{w_i}\right)^{\rho_i^j -1}
$$

\begin{equation}\label{eq:12}
    = \rho_i^j \frac{q_i^j}{(A_i^j)^{1-\rho_i^j}} \left[ w_i^{\beta_i^j(1-\rho_i^j)} \left(P_i^j \right)^{(1-\beta_i^j)(1-\rho_i^j)} [t_i(1-\epsilon_i^j)]^{\rho_i^j -1} \right]
\end{equation}

 Taking the derivative of the total cost function with respect to $q_i^j$, computed at the optimal input combination, provides the marginal cost of the input bundle, given by:

$$
    mc_i^j \equiv \frac{\partial TC_i^j}{\partial q_i^j} = \frac{\partial (w_il_i^j + P_i^jM_i^j +t_ie_i^j)}{\partial q_i^j}  =
$$    
$$
    = \left[ \beta_i^j(1-\rho_i^j) + (1-\beta_i^j)(1-\rho_i^j) + \rho_i^j   \right] \left[ \frac{1}{(A_i^j)^{1-\rho_i^j}} \left[ w_i^{\beta_i^j} (P_i^j)^{1-\beta_i^j}\right]^{1- \rho_i^j} [t_i(1-\epsilon_i^j)]^{\rho_i^j} \right] =
$$

\begin{equation*}
    =  \frac{1}{(A_i^j)^{1-\rho_i^j}} \left[ w_i^{\beta_i^j} (P_i^j)^{1-\beta_i^j}\right]^{1- \rho_i^j} [t_i(1-\epsilon_i^j)]^{\rho_i^j}
\end{equation*}.

\paragraph{Steady state and normalisation} We define the steady state as the combination of equilibrium prices and quantities in an undistorted economy with no wedges, where $ (1-\epsilon_n^k)^{\rho_n^k}(A_n^k)^{-(1-\rho_n^k)} = 1 \quad \forall i,n \in \mathcal{N} \ j,k \in \mathcal{J}$. From Lemma A.2. and the pricing rule we have: 

$$p_{i}^{j} =  mc_i^j = \left[(A_i^j)^{-(1- \rho_i^j)} w_i^{\beta_i^j(1-\rho_i^j)}  (P_i^j)^{(1-\beta_i^j)(1-\rho_i^j)} [t_i(1-\epsilon_i^j)]^{\rho_i^j} \right]$$

 Taking its logs and evaluating it at the steady-state:

$$ \log p_{i}^{j} = \beta_i^j(1-\rho_i^j)\log w_i + (1-\beta_i^j)(1-\rho_i^j) \log   \left( \sum_{n \in \mathcal{N}} \sum_{k \in \mathcal{J}} \iota_{ni}^{kj} (p_{n}^{k})^{1-\theta}\right)^{\frac{1}{1-\theta}} + \rho_i^j \log t_i $$

 And the system of equations is solved for $p_{i}^{j} = w_i = t_i = 1 \quad \forall i,j$. \\

\paragraph{Comparative Statics}

We provide comparative statics that offer a first-order approximation of the effect of the CBAM on trade across sectors, welfare, and emissions. To account for the CBAM endogeneity, we proceed as follows. We define the steady state of the economy as one with no trade wedges. We write the tariff component of the price wedge on sectoral trade as  $\tilde{\tau}_{ni}^{kj}= (1+ \kappa_{ni}^{kj} + h_{ni}^{kj})$, where $h_{ni}^{kj}$ denotes the CBAM imposed by country $i$ on imports from sector $k$ in country $n$. In the steady state, both $\kappa_{ni}^{kj}$ and $h_{ni}^{kj}$ are equal to zero. To capture the marginal impact of introducing the CBAM, we take the derivative of the relevant outcome variable with respect to $h_{ni}^{kj}$ evaluated at the steady state. When constructing the first-order approximation of the CBAM’s effect, we incorporate the feedback that the additional tariff has on emission prices -- and consequently on the actual CBAM level -- using the definition of the CBAM tariff in \eqref{eq:CBAM_definition}.\\

 We start by establishing two important lemmas that underpin the analysis. These results provide the analytical structure necessary to understand how the introduction of the additional carbon tariff propagates and affects intermediate input prices and how the CBAM tariff adjusts endogenously to shifts in carbon prices. 

\begin{lemma} \textbf{(Prices)} For a shock to the trade costs, the change in the price of intermediate inputs in the steady state is:
    \begin{equation}\label{derPRICES}
     \frac{\partial \log \mathbf{p}}{\partial h_{ls}^{qr}} = \left(\mathbf{I} - \bm{\gamma} \mathbf{\Pi'} \right)^{-1} \left[\bm{\beta}(\mathbf{I} - \bm{\rho}) \frac{\partial \log \mathbf{w}}{\partial h_{ls}^{qr}} + \bm{\rho} \frac{\partial \log \mathbf{t}}{\partial h_{ls}^{qr}} + \bm{\gamma} diag^{-1}\left( \mathbf{\Pi'} \mathbf{e}_{2(l-1)+q} \mathbf{e'}_{2(s-1)+r}\right)\right]
    \end{equation}
\end{lemma}

 \textit{Proof.} Consider first the price of intermediate inputs and the pricing rule. In equilibrium:

$$ p_{n}^{k} = mc_n^k = (A_n^k)^{-(1- \rho_n^k)} w_n^{\beta_n^k(1-\rho_n^k)}  (P_n^k)^{(1-\beta_n^k)(1-\rho_n^k)} [t_n(1-\epsilon_n^k)^{\rho_n^k}] $$

 Taking its logarithm:

\begin{equation*}
    \log p_n^k = -(1-\rho_n^k) \log A_n^k + \beta_n^k(1-\rho_n^k) \log w_n + (1-\beta_n^k)(1-\rho_n^k) \log P_n^k + \rho_n^k \log [t_n(1-\epsilon_n^k)] 
\end{equation*}

 Given our modelling assumptions, the CBAM enters as an additional tariff, as shown in (\ref{eqCBAM}). Thus, we now consider a marginal increase in trade wedges, represented by an arbitrarily small positive $h_{ls}^{qr}$, and derive the percentage change in intermediate input prices in response to the shock. Differentiating with respect to $ h_{ls}^{qr}$ and evaluating it at the steady-state:

\begin{equation}
    \frac{\partial \log p_n^k}{\partial h_{ls}^{qr}} = \beta_n^k(1-\rho_n^k) \frac{\partial \log w_n}{\partial h_{ls}^{qr}} + (1-\beta_n^k)(1-\rho_n^k) \frac{\partial \log P_n^k}{\partial h_{ls}^{qr}} + \rho_n^k \frac{\partial \log t_n}{\partial h_{ls}^{qr}} 
\end{equation}

 where from the definition of the price index $P_n^k = \left( \sum_{m \in \mathcal{N}} \sum_{h \in \mathcal{J}} \iota_{mn}^{hk} (p_{mn}^{hk})^{1-\theta} \right)^{\frac{1}{1-\theta}}$, we have:\\

$$\frac{\partial\log P_n^k}{\partial h_{ls}^{qr}} = \frac{1}{\sum_{m \in \mathcal{N}} \sum_{h \in \mathcal{J}} \iota_{mn}^{hk} (p_{mn}^{hk})^{1-\theta}} \sum_{m \in \mathcal{N}} \sum_{h \in \mathcal{J}} \iota_{mn}^{hk} (p_{mn}^{hk})^{1-\theta} \frac{\partial \log p_{mn}^{hk}}{ \partial h_{ls}^{qr}}$$\\

 which in steady-state becomes:

$$\frac{\partial\log P_n^k}{\partial h_{ls}^{qr}} = \sum_{m \in \mathcal{N}} \sum_{h \in \mathcal{J}} \iota_{mn}^{hk} \frac{\partial \log p_{m}^{h}}{\partial h_{ls}^{qr}} +  \sum_{m \in \mathcal{N}} \sum_{h \in \mathcal{J}} \iota_{mn}^{hk} \frac{\partial \log \tau_{mn}^{hk}}{\partial h_{ls}^{qr}} $$

 where:
\begin{equation}\label{derTARIFFS}
    \frac{\partial \log \tau_{mn}^{hk}}{\partial h_{ls}^{qr}} =  \frac{\partial h_{mn}^{hk}}{\partial h_{ls}^{qr}} = 1_{ls = mn} \cdot 1_{qr = hk}
\end{equation}

 Defining $\mathbf{\Pi} \in \mathcal{M}(NJ, NJ)$ the matrix with entries $\iota_{ni}^{kj}$ and with $\mathbf{p} \in \mathbb{R}^{NJ}$ the vector of prices $p_i^j$, in matrix notation we have:
\begin{equation}\label{PI}
 \frac{\partial\log P_n^k}{\partial h_{ls}^{qr}} = \mathbf{e'}_{2(n-1)+k} \mathbf{\Pi'} \frac{\partial \log \mathbf{p}}{\partial h_{ls}^{qr}} + \mathbf{e'}_{2(n-1)+k} diag^{-1} \left( \mathbf{\Pi'} \mathbf{e}_{2(l-1)+q} \mathbf{e'}_{2(s-1)+r}\right)
\end{equation}

 and the derivative of the vector of log prices can be rewritten as:\\

$$ 
\frac{\partial \log \mathbf{p}}{\partial h_{ls}^{qr}} = \bm{\beta} (\mathbf{I} - \bm{\rho}) \frac{\partial \log \mathbf{w}}{\partial h_{ls}^{qr}} +  \bm{\rho} \frac{\partial \log \mathbf{t}}{\partial h_{ls}^{qr}} + \bm{\gamma} \mathbf{\Pi'} \frac{\partial \log \mathbf{p}}{\partial h_{ls}^{qr}} + \bm{\gamma} diag^{-1}\left( \mathbf{\Pi'} \mathbf{e}_{2(l-1)+q} \mathbf{e'}_{2(s-1)+r}\right)
$$\\

 where $ \bm{\rho} = diag(\rho_1^1, \rho_1^2, ..., \rho_N^J)' \in \mathcal{M}(NJ, NJ) $, $ \bm{\beta} = diag(\beta_1^1, \beta_1^2, ..., \beta_N^J)' \in \mathcal{M}(NJ,NJ) $, $ \bm{\gamma} = (\mathbf{I} - \bm{\beta})(\mathbf{I} -\bm{\rho}) \in \mathcal{M}(NJ,NJ)$, $\mathbf{w} = (w_1, ..., w_N)' \otimes \mathbf{1}_{J} \in \mathbb{R}^{NJ}$ and $\mathbf{t} = (t_1,..., t_N)' \otimes \mathbf{1}_{J}$\footnote{$\otimes$ denotes the Kronecker product.}. Thus: 
$$ \frac{\partial \log \mathbf{p}}{\partial h_{ls}^{qr}} = \left(\mathbf{I} - \bm{\gamma} \mathbf{\Pi'} \right)^{-1} \left[\bm{\beta}(\mathbf{I} - \bm{\rho}) \frac{\partial \log \mathbf{w}}{\partial h_{ls}^{qr}} + \bm{\rho} \frac{\partial \log \mathbf{t}}{\partial h_{ls}^{qr}} + \bm{\gamma} diag^{-1}\left( \mathbf{\Pi'} \mathbf{e}_{2(l-1)+q} \mathbf{e'}_{2(s-1)+r}\right)\right] $$

 And in the special case where $\beta_i^j = \beta, \ \rho_i^j = \rho \ \forall i,j$:

$$ \frac{\partial \log \mathbf{p}}{\partial h_{ls}^{qr}} = \left(\mathbf{I} - \gamma \mathbf{\Pi'} \right)^{-1} \left[ \beta(1-\rho) \frac{\partial \log \mathbf{w}}{\partial h_{ls}^{qr}} +\rho \frac{\partial \log \mathbf{t}}{\partial h_{ls}^{qr}} + \gamma  diag^{-1}\left( \mathbf{\Pi'} \mathbf{e}_{2(l-1)+q} \mathbf{e'}_{2(s-1)+r}\right)\right] $$\

\begin{lemma} \textbf{(CBAM)}
    For a shock to the trade cost, the first-order approximation around the steady-state of the change in the CBAM wedge applied on imports from sector $q$, country $l$ and directed towards sector $r$ in country $s$ is:
    \begin{equation}\label{derCBAM}
        d CBAM_{ls}^{qr} =  (\rho_l^q)^2 \left( \frac{\partial \log t_s}{\partial h_{ls}^{qr}} - \frac{\partial \log t_l}{\partial h_{ls}^{qr}} \right) 
        \end{equation}
\end{lemma}

 \textit{Proof.} Consider the change in the CBAM wedge, focusing on the more general case in which $CBAM_{ls}^{qr} = \rho_l^q \frac{t_s}{t_l}$. To a first-order: 

$$
d CBAM_{ls}^{qr} = d \left( \rho_l^q \frac{t_s}{t_l} \right) = \frac{\partial \left( \rho_l^q \frac{t_s}{t_l} \right)}{\partial h_{ls}^{qr}} \left( \rho_l^q \frac{t_s}{t_l} + d \left( \rho_l^q \frac{t_s}{t_l} \right) \right) = 
$$

$$
= \frac{{ \frac{  \partial \left( \rho_l^q \frac{t_s}{t_l} \right)}{\partial h_{ls}^{qr}} } \rho_l^q \frac{t_s}{t_l} }{1- \frac{\partial \left( \rho_l^q \frac{t_s}{t_l} \right)}{\partial h_{ls}^{qr}} }
$$
 which, for small shocks, is well approximated by:

$$ 
d CBAM_{ls}^{qr}  =  \frac{\partial \left( \rho_l^q \frac{t_s}{t_l} \right)}{\partial h_{ls}^{qr}} \rho_l^q \frac{t_s}{t_l} = \rho_l^q \frac{t_s}{t_l} \left( \rho_l^q \frac{(\partial t_s/ \partial h_{ls}^{qr})t_l - (\partial t_l/ \partial h_{ls}^{qr})t_s}{t_l^2} \right)
$$
 Evaluating it at the steady-state:

$$
d CBAM_{ls}^{qr}  =  (\rho_l^q)^2 \left( \frac{\partial \log t_s}{\partial h_{ls}^{qr}} - \frac{\partial \log t_l}{\partial h_{ls}^{qr}} \right) $$

 With these components in place, we can now characterise the adjustment in cost shares, welfare and emissions embodied in imports resulting from the introduction of the CBAM. 

\begin{proposition} \textbf{(Cost Shares)}  \textit{The first-order approximation around the steady state of the change in the cost shares following the introduction of the CBAM is given by:}

$$d \log \tilde{\omega}_{ni}^{kj} = (1- \theta) \left[ (\rho_n^k + dCBAM_{ni}^{kj}) + (1-\beta_n^k) (1-\rho_n^k)  \sum_{l \in \mathcal{N}} \sum_{s \in \mathcal{N}} \sum_{q \in \mathcal{J}} \sum_{r \in \mathcal{J}} \psi_{ns}^{kr} \iota_{ls}^{qr} (\rho_l^q + dCBAM_{ls}^{qr}) \right. +  $$

$$ - \left. \sum_{l \in \mathcal{N}} \sum_{s \in \mathcal{N}} \sum_{q \in \mathcal{J}} \sum_{r \in \mathcal{J}} \psi_{is}^{jr} \iota_{ls}^{qr} (\rho_l^q + dCBAM_{ls}^{qr}) \right] + $$
$$ + (1 - \theta) \sum_{l \in \mathcal{N}} \sum_{s \in \mathcal{N}} \sum_{q \in \mathcal{J}} \sum_{r \in \mathcal{J}} \left[ \mathbf{e'}_{2(n-1)+k} (\mathbf{I} - \gamma\mathbf{\Pi'})^{-1} \left( \beta (1-\rho) \frac{\partial \log \mathbf{w}}{\partial h_{ls}^{qr}} + \rho \frac{\partial \log \mathbf{t}}{\partial h_{ls}^{qr}} \right) - \right. $$
\begin{equation}\label{derCOSTSHARES}
    \left. \mathbf{e'}_{2(i-1)+j} (\mathbf{I} - \gamma\mathbf{\Pi'})^{-1} \left( \beta (1-\rho) \frac{\partial \log \mathbf{w}}{\partial h_{ls}^{qr}} + \rho \frac{\partial \log \mathbf{t}}{\partial h_{ls}^{qr}}  \right) \right] (\rho_l^q + dCBAM_{ls}^{qr})
\end{equation}

 where $dCBAM_{ls}^{qr}$ is given by (\ref{derCBAM}).
\end{proposition}

 \textit{Proof.} Consider the cost shares $\tilde{\omega}_{ni}^{kj}$. In equilibrium, from (\ref{eqINTINPUTS}) and the pricing rule we have:

$$\tilde{\omega}_{ni}^{kj} = \iota_{ni}^{kj} (p_n^k \tau_{ni}^{kj})^{1-\theta}(P_i^j)^{-(1-\theta)} $$

 Taking its logarithm:

$$ \log \tilde{\omega}_{ni}^{kj} = \log \iota_{ni}^{kj} + (1-\theta) \left[ \log p_n^k + \log \tau_{ni}^{kj} - log P_i^j \right]$$

 and then differentiating with respect to $h_{ls}^{qr}$ and evaluating it at the steady-state:

$$ \frac{\partial \log \tilde{\omega}_{ni}^{kj}}{\partial h_{ls}^{qr}} = (1- \theta) \left[ \frac{\partial \log p_n^k}{\partial h_{ls}^{qr}} + \frac{\partial \log \tau_{ni}^{kj}}{\partial h_{ls}^{qr}} - \frac{\partial \log P_i^j}{\partial h_{ls}^{qr}} \right] = $$

$$
= (1- \theta) \left[  1_{ls=ni} \cdot 1_{qr=kj} 
 + \beta_n^k (1-\rho_n^k) \frac{\partial \log w_n}{\partial h_{ls}^{qr}} + \rho_n^k \frac{\partial \log t_n}{\partial h_{ls}^{qr}} + \right. $$
$$ + (1-\beta_n^k) (1-\rho_n^k) \left. \mathbf{e'}_{2(n-1)+k} \left( \mathbf{\Pi'} \frac{\partial \log \mathbf{p}}{\partial h_{ls}^{qr}} + diag^{-1} \left( \mathbf{\Pi'}  \mathbf{e}_{2(l-1)+q} \mathbf{e'}_{2(s-1)+r}  \right)\right) + \right. $$
 $$ \left. -\mathbf{e'}_{2(i-1)+j} \left( \mathbf{\Pi'} \frac{\partial \log \mathbf{p}}{\partial h_{ls}^{qr}} + diag^{-1} \left( \mathbf{\Pi'}  \mathbf{e}_{2(l-1)+q} \mathbf{e'}_{2(s-1)+r} \right) \right) \right]
$$

 where $\frac{d \log \mathbf{p}}{d \log h_{ls}^{qr}} $ is given by (\ref{derPRICES}). Thus: 

$$ \frac{\partial \log \tilde{\omega}_{ni}^{kj}}{\partial h_{ls}^{qr}}= (1-\theta) \left\{ 1_{ls=ni} \cdot 1_{qr=kj} + \mathbf{e'}_{2(n-1) + k} \left[ \bm{\beta} (\mathbf{I} - \bm{\rho}) \frac{\partial \log \mathbf{w}}{\partial h_{ls}^{qr}} +  \bm{\rho} \frac{\partial \log \mathbf{t}}{\partial h_{ls}^{qr}} + \right. \right. $$

$$ \left. + \bm{\gamma} \mathbf{\Pi'} (\mathbf{I} - \bm{\gamma} \mathbf{\Pi'})^{-1} \left( \bm{\beta} (\mathbf{I}-\bm{\rho}) \frac{\partial \log \mathbf{w}}{\partial h_{ls}^{qr}} +  \bm{\rho} \frac{\partial \log \mathbf{t}}{\partial h_{ls}^{qr}} + \bm{\gamma} diag^{-1} (\mathbf{\Pi'} \mathbf{e}_{2(l-1)+q} \mathbf{e'}_{2(s-1)+r}) \right) \right. $$

$$ + \bm{\gamma} diag^{-1} (\mathbf{\Pi'} \mathbf{e}_{2(l-1)+q} \mathbf{e'}_{2(s-1)+r}) \bigg] +  $$

$$ \left. - \mathbf{e'}_{2(i-1) + j} \left[ \mathbf{\Pi'(\mathbf{I} - \bm{\gamma} \mathbf{\Pi'} )^{-1}} \left( \bm{\beta} (\mathbf{I}-\bm{\rho}) \frac{\partial \log \mathbf{w}}{\partial h_{ls}^{qr}} +  \bm{\rho} \frac{\partial \log \mathbf{t}}{\partial h_{ls}^{qr}} + \bm{\gamma} diag^{-1} (\mathbf{\Pi'} \mathbf{e}_{2(l-1)+q} \mathbf{e'}_{2(s-1)+r}) \right) \right. \right. $$

$$ + diag^{-1} (\mathbf{\Pi'} \mathbf{e}_{2(l-1)+q} \mathbf{e'}_{2(s-1)+r}) \bigg] \bigg\}$$\\

 And using the fact that $\mathbf{I} + \bm{\gamma} \mathbf{\Pi'}(\mathbf{I} - \bm{\gamma} \mathbf{\Pi'})^{-1} = (\mathbf{I} -\bm{\gamma} \mathbf{\Pi'})^{-1}$: 

$$
= (1-\theta) \left\{ 1_{ls=ni} \cdot 1_{qr=kj} + \mathbf{e'}_{2(n-1) + k} \left[ (\mathbf{I} - \bm{\gamma} \mathbf{\Pi'})^{-1} \left( \bm{\beta} (\mathbf{I}-\bm{\rho}) \frac{\partial \log \mathbf{w}}{\partial h_{ls}^{qr}} +  \bm{\rho} \frac{\partial \log \mathbf{t}}{\partial h_{ls}^{qr}} \right) \right] \right. + 
$$

$$
- \mathbf{e'}_{2(i-1) + j} \left[ \mathbf{\Pi'(\mathbf{I} - \bm{\gamma} \mathbf{\Pi'} )^{-1}} \left( \bm{\beta} (\mathbf{I}-\bm{\rho}) \frac{\partial \log \mathbf{w}}{\partial h_{ls}^{qr}} +  \bm{\rho} \frac{\partial \log \mathbf{t}}{\partial h_{ls}^{qr}} \right) \right] +
$$

$$
+ \mathbf{e'}_{2(n-1) + k} (\mathbf{I} - \bm{\gamma} \mathbf{\Pi'} )^{-1} \bm{\gamma} diag^{-1} (\mathbf{\Pi'} \mathbf{e}_{2(l-1)+q} \mathbf{e'}_{2(s-1)+r}) 
$$

$$
- \mathbf{e'}_{2(i-1) + j}  \left[ \mathbf{I} + \mathbf{\Pi'} (\mathbf{I} - \bm{\gamma} \mathbf{\Pi'} )^{-1} \bm{\gamma} \right]  diag^{-1} (\mathbf{\Pi'} \mathbf{e}_{2(l-1)+q} \mathbf{e'}_{2(s-1)+r}) \Big\} =
$$

$$
= (1-\theta) \left\{ 1_{ls=ni} \cdot 1_{qr=kj} + \mathbf{e'}_{2(n-1) + k} \left[ (\mathbf{I} - \bm{\gamma} \mathbf{\Pi'})^{-1} \left( \bm{\beta} (\mathbf{I}-\bm{\rho}) \frac{\partial \log \mathbf{w}}{\partial h_{ls}^{qr}} +  \bm{\rho} \frac{\partial \log \mathbf{t}}{\partial h_{ls}^{qr}} \right) \right] \right. + 
$$

$$
- \mathbf{e}_{2(i-1) + j} \left[ \mathbf{\Pi'(\mathbf{I} - \bm{\gamma} \mathbf{\Pi'} )^{-1}} \left( \bm{\beta} (\mathbf{I}-\bm{\rho}) \frac{\partial \log \mathbf{w}}{\partial h_{ls}^{qr}} +  \bm{\rho} \frac{\partial \log \mathbf{t}}{\partial h_{ls}^{qr}} \right) \right] +
$$

\begin{equation}\label{eqCOSTSHARES}
    \left. + (1-\beta_r^s)(1-\rho_r^s) \psi_{ns}^{kr} \iota_{ls}^{qr} - 1_{ls=ni} \cdot 1_{qr=kj} \cdot \iota_{ls}^{qr} - \sum_{m \in \mathcal{N}} \sum_{h \in \mathcal{J}} (1-\beta_r^s)(1-\rho_r^s) \iota_{mi}^{hj} \psi_{ms}^{hr} \iota_{ls}^{qr} \right\}
\end{equation}\\

 where with $\psi_{in}^{jk}$ we denote the entries of the cost-based Leontief inverse $(\mathbf{I} - \bm{\gamma} \mathbf{\Pi'} )^{-1} \in \mathcal{M}(NJ,NJ)$. Moreover, since $\omega_{ni}^{kj} = (1-\beta_i^j)(1-\rho_i^j) \frac{\tilde{\omega}_{ni}^{kj}}{\tilde{\tau}_{ni}^{kj}}$, we have:

\begin{equation}\label{derREVENUESHARES}
    \frac{\partial \log \omega_{ni}^{kj}}{\partial h_{ls}^{qr}} =  \frac{\partial \log \tilde{\omega}_{ni}^{kj}}{\partial h_{ls}^{qr}} - \frac{\partial \log \tilde{\tau}_{ni}^{kj}}{\partial h_{ls}^{qr}} = \frac{\partial \log \tilde{\omega}_{ni}^{kj}}{\partial h_{ls}^{qr}} - 1_{ni=ls} \cdot 1_{kj=qr} 
\end{equation}

 In the case in which $ \beta_i^j = \beta, \ \rho_i^j = \rho \ \forall i, j$:

$$\frac{\partial \log \tilde{\omega}_{ni}^{kj}}{\partial h_{ls}^{qr}} = (1-\theta) \left\{ 1_{ls=ni} \cdot 1_{qr=kj} + \mathbf{e'}_{2(n-1) + k} \left[ (\mathbf{I} - \gamma \mathbf{\Pi'})^{-1} \left( \beta (1-\rho) \frac{\partial \log \mathbf{w}}{\partial h_{ls}^{qr}} +  \rho \frac{\partial \log \mathbf{t}}{\partial h_{ls}^{qr}} \right) \right] \right. +
$$

$$
- \mathbf{e}_{2(i-1) + j} \left[ \mathbf{\Pi'(\mathbf{I} - \gamma \mathbf{\Pi'} )^{-1}} \left( \beta (1-\rho) \frac{\partial \log \mathbf{w}}{\partial h_{ls}^{qr}} +  \rho \frac{\partial \log \mathbf{t}}{\partial h_{ls}^{qr}} \right) \right] +
$$

$$ \left. + (1-\beta)(1-\rho) \psi_{ns}^{kr} \iota_{ls}^{qr} - \psi_{is}^{jr} \iota_{ls}^{qr}  \right\}$$

 Now, the percentage change of cost shares following the introduction of the marginal increase in trade wedges can be approximated by the following first-order Taylor approximation around the steady-state:\

$$d \log \tilde{\omega}_{ni}^{kj} = \sum_{l \in \mathcal{N}} \sum_{s \in \mathcal{N}} \sum_{q \in \mathcal{J}} \sum_{r \in \mathcal{J}} \frac{\partial \log \tilde{\omega}_{ni}^{kj}}{\partial h_{ls}^{qr}} d  h_{ls}^{qr} = $$ 

$$ = \sum_{l \in \mathcal{N}} \sum_{s \in \mathcal{N}} \sum_{q \in \mathcal{J}} \sum_{r \in \mathcal{J}} \frac{\partial \log \tilde{\omega}_{ni}^{kj}}{\partial h_{ls}^{qr}} \left( \rho_l^q +  dCBAM_{ls}^{qr} \right)$$\

 where $\frac{\partial \log \tilde{\omega}_{ni}^{kj}}{\partial h_{ls}^{qr}}$ is given by (\ref{derCOSTSHARES}) and $dCBAM_{ls}^{qr}$ by (\ref{derCBAM}). However, for our purposes, we consider the simplified version in which a unique country $i$ introduces the CBAM on a special good $k$ produced in country $n$ and factor shares are uniform across countries and sectors. We can rewrite the above expression as:

$$d \log \tilde{\omega}_{ni}^{kj} = (1- \theta) \left[ (\rho + dCBAM_{ni}^{kj}) + (1-\beta) (1-\rho) \sum_{r \in \mathcal{J}} \left(  \psi_{ni}^{kr} - \psi_{ii}^{jr}   \right) \iota_{ni}^{kr}(\rho + dCBAM_{ni}^{kr})  \right] + $$
\normalsize
$$ +(1-\theta) \left[ \mathbf{e'}_{2(n-1)+k} (1 - \gamma \mathbf{\Pi'})^{-1}  \left( \beta (1- \rho) \frac{\partial \log \mathbf{w}}{\partial \mathbf{h}} + \rho \frac{\partial \log \mathbf{t}}{\partial \mathbf{h}}  \right) + \right. $$
$$ \left. -\mathbf{e'}_{2(i-1)+j} (1 - \gamma \mathbf{\Pi'})^{-1} \left( \beta (1- \rho) \frac{\partial \log \mathbf{w}}{\partial \mathbf{h}} + \rho \frac{\partial \log \mathbf{t}}{\partial \mathbf{h}} \right) \right] (\bm{\rho} + d \mathbf{CBAM}) $$

 where $\mathbf{h} = (h_{ni}^{k1}, h_{ni}^{k2}, ..., h_{ni}^{kN}) \in \mathbb{R}^{NJ}$ and $ \mathbf{CBAM} = (CBAM_{ni}^{k1}, CBAM_{ni}^{k2}, ...,CBAM_{ni}^{kN}) \in \mathbb{R}^{NJ}$, with $dCBAM_{ni}^{kj}$ given by (\ref{derCBAM}). All derivatives are evaluated at the steady state, where $\mathbf{h} =\mathbf{0}$.\\

\begin{proposition} (\textbf{Gross National Expenditure})
The first-order approximation of the change in real GNE of country $i$ following the introduction of the CBAM, around the steady-state is given by:
\small{
$$d \log W_i =  \sum_{l \in \mathcal{N}} \sum_{s \in \mathcal{N}} \sum_{q \in \mathcal{J}} \sum_{r \in \mathcal{J}} \left[  \frac{w_i \overline{L}_i}{I_i} \frac{\partial \log w_i}{\partial h_{ls}^{qr}} + \frac{t_i E_i}{I_i} \left( \mathbf{1}_{\mathcal{N}_{c(i)}} \frac{ \partial \log t_i}{\partial h_{ls}^{qr}}  + \mathbf{1}_{\mathcal{N}_{nc(i)}} \frac{ \partial \log E_i}{\partial h_{ls}^{qr}} \right) \right] (\rho_l^q + dCBAM_{ls}^{qr}) $$}
\normalsize
$$- \sum_{l \in \mathcal{N}} \sum_{s \in \mathcal{N}} \sum_{q \in \mathcal{J}} \sum_{r \in \mathcal{J}} (\mathbf{e}_i \otimes \mathbf{1}_{J})' \bm{\chi} \frac{\partial \log \mathbf{p}}{\partial h_{ls}^{qr}} (\rho_l^q + dCBAM_{ls}^{qr}) + $$

\begin{equation}
      +  (1-\beta_i^j) (1-\rho_i^j) \lambda_i^j \sum_{l \in \mathcal{N}} \sum_{q \in \mathcal{J}} \tilde{\omega}_{li}^{qj} (\rho_l^q + dCBAM_{ls}^{qr}) 
 \end{equation}
 where $\mathbf{e}_i \in \mathbb{R}^N$ is th \textit{i}-th standard basis vector, $\bm{\chi} = diag^{-1}(\chi_1^1, ..., \chi_N^J)' \in \mathcal{M}(NJ\times NJ)$ is the diagonal matrix of consumption shares $\chi_i^j$, that in equilibrium are equal to $\nu_i^j$, $\frac{\partial \log \mathbf{p}}{\partial h_{ls}^{qr}}$ is given by (\ref{derPRICES}) and $dCBAM_{ls}^{qr}$ is given by (\ref{derCBAM}).
\end{proposition}
\vspace{3mm}

 \textit{Proof.} We recall the definition of each country's nominal Gross National Expenditure (GNE) given by $GNE_i = \sum_{j \in \mathcal{J}} p_i^j c_i^j$ and define $\nu_i^j = p_i^jc_i^j/GNE_i$. Thus, taking derivatives with respect to $h_{ls}^{qr}$, the percentage change in nominal Gross National Expenditure of country $i$ is equal to:

\begin{equation}
    \frac{\partial \log GNE_i}{\partial h_{ni}^{kj}}  = \underbrace{\sum_{j \in \mathcal{J}} \nu_i^j \frac{ \partial \log p_i^j}{\partial h_{ni}^{kj}}}_{\text{price effect}} + \underbrace{\sum_{j \in \mathcal{J}} \nu_i^j  \frac{\partial \log c_i^j}{\partial h_{ni}^{kj}}}_{\text{real effect}} \equiv  \frac{ \partial \log P_i}{\partial h_{ni}^{kj}} + \frac{\partial \log W_i}{\partial h_{ni}^{kj}} 
\end{equation}

 where $\frac{\partial \log p_i^j}{\partial h_{ls}^{qr}}$ is equal to the ($2(i-1) + j$)\textit{th} element of the vector $\frac{\partial \log \mathbf{p}}{\partial h_{ls}^{qr}}$ in (\ref{derPRICES}).\\

 Given the assumption of locally non-satiated preferences, total expenditure on final goods of country $i$ is equal to the national income  $I_i$, coinciding with:

$$
I_i = w_i \overline{L}_i + t_i E_i + \sum_{j \in \mathcal{J}} (1-\beta_i^j) (1-\rho_i^j) p_i^j q_i^j \sum_{n \in \mathcal{N}}\sum_{k \in \mathcal{J} }\left( \tilde{\tau}_{ni}^{kj} - 1\right) \tilde{\omega}_{ni}^{kj} + \overline{D_i} 
$$

 Differentiating and using the fact that $d \log x =d x/x$, we have:

$$ \frac{\partial \log I_i}{\partial h_{ls}^{qr}}  =  \frac{w_i \overline{L}_i}{I_i} \frac{\partial \log w_i}{\partial h_{ls}^{qr}} + \frac{t_i E_i}{I_i} \left( \mathbf{1}_{\mathcal{N}_{c(i)}} \frac{ \partial \log t_i}{\partial h_{ls}^{qr}} + \mathbf{1}_{\mathcal{N}_{nc(i)}} \frac{ \partial \log E_i}{\partial h_{ls}^{qr}} \right) + \frac{1}{I_i} \frac{\partial R_i}{\partial h_{ls}^{qr}}
 $$

 where $R_i =  \sum_{j \in \mathcal{J}} (1-\beta_i^j) (1-\rho_i^j) \lambda_i^j GNE \sum_{n \in \mathcal{N}}\sum_{k \in \mathcal{J} }\left( \tilde{\tau}_{ni}^{kj} - 1\right) \tilde{\omega}_{ni}^{kj} $ and in steady-state:

$$ \frac{\partial R_i}{\partial h_{ls}^{qr}} = \sum_{j \in \mathcal{J}} (1-\beta_i^j) (1-\rho_i^j) \lambda_i^j \sum_{n \in \mathcal{N}}\sum_{k \in \mathcal{J} } \frac{ \partial \log \tilde{\tau}_{ni}^{kj}}{\partial h_{ls}^{qr}} \tilde{\omega}_{ni}^{kj} = $$

$$ = 1_{ls=ni} \cdot 1_{qr=kj} \cdot 
 \sum_{j \in \mathcal{J}} (1-\beta_i^j) (1-\rho_i^j) \lambda_i^j \tilde{\omega}_{ni}^{kj} $$

 Since, in equilibrium, $GNE_i = I_i$ the percentage change in nominal Gross National Expenditure of county $i$ is given by:

$$ \frac{\partial \log GNE_i}{\partial h_{ls}^{qr}} = \left[  \frac{w_i \overline{L}_i}{I_i} \frac{\partial \log w_i}{\partial h_{ls}^{qr}} + \frac{t_i E_i}{I_i} \left( \mathbf{1}_{\mathcal{N}_{c(i)}} \frac{ \partial \log t_i}{\partial h_{ls}^{qr}} + \mathbf{1}_{\mathcal{N}_{nc(i)}} \frac{ \partial \log E_i}{\partial h_{ls}^{qr}} \right)+ \right.$$
\begin{equation}
 + \left. 1_{ls=ni} \cdot 1_{qr=kj} \cdot \sum_{j \in \mathcal{J}} (1-\beta_i^j) (1-\rho_i^j) \lambda_i^j \tilde{\omega}_{ni}^{kj} \right]
\end{equation}

 And the corresponding change in real GNE, reflecting changes in welfare, is given by:

$$ \frac{\partial \log W_i}{\partial h_{ls}^{qr}} = \frac{\partial \log GNE_i}{\partial h_{ls}^{qr}} - \frac{\partial \log P_i}{\partial h_{ls}^{qr}} = $$

$$=  \left[  \frac{w_i \overline{L}_i}{I_i} \frac{\partial \log w_i}{\partial h_{ls}^{qr}} + \frac{t_i E_i}{I_i} \left( \mathbf{1}_{\mathcal{N}_{c(i)}} \frac{ \partial \log t_i}{\partial h_{ls}^{qr}} + \mathbf{1}_{\mathcal{N}_{nc(i)}} \frac{ \partial \log E_i}{\partial h_{ls}^{qr}} \right) + \right. $$
$$ + \left. 1_{ls=ni} \cdot 1_{qr=kj} \cdot \sum_{r \in \mathcal{J}} (1-\beta_s^r) (1-\rho_s^r) \lambda_s^r \tilde{\omega}_{ls}^{qr} \right] + $$

\begin{equation}\label{derWELFARE}
    - (\mathbf{e}_i \otimes \mathbf{1}_{J})' \bm{\chi} \frac{\partial \log \mathbf{p}}{\partial h_{ls}^{qr}}
\end{equation}

 where $\mathbf{e}_i \in \mathbb{R}^N$ is th \textit{i}-th standard basis vector, $\bm{\chi} = diag^{-1}(\chi_1^1, ..., \chi_N^J)' \in \mathcal{M}(NJ\times NJ)$ is the diagonal matrix of consumption shares $\chi_i^j$, that in equilibrium are equal to $\nu_i^j$ and $\frac{\partial \log \mathbf{p}}{\partial h_{ls}^{qr}}$ is given by (\ref{derPRICES}).\\

 Finally, the first-order approximations of the percentage change in real Gross National Expenditure following the marginal increase in trade wedges is given by: 

$$d \log W_i = \sum_{l \in \mathcal{N}} \sum_{s \in \mathcal{N}} \sum_{q \in \mathcal{J}} \sum_{r \in \mathcal{J}} \frac{\partial \log W_i}{\partial h_{ls}^{qr}} d h_{ls}^{qr} = $$

$$
\sum_{l \in \mathcal{N}} \sum_{s \in \mathcal{N}} \sum_{q \in \mathcal{J}} \sum_{r \in \mathcal{J}} \frac{\partial \log W_i}{\partial h_{ls}^{qr}} \left( \rho_l^q +  dCBAM_{ls}^{qr} \right)
$$

 where $dCBAM_{ls}^{qr} $ is given by (\ref{derCBAM}) and $\frac{\partial \log W_i}{\partial h_{ls}^{qr}}$ by (\ref{derWELFARE}). In order to gain a better insight into the mechanisms driving the change, we consider the special case in which a unique country $i$ introduces the CBAM on a specific good $k$ produced in country $n$ and we set factor shares equal across countries and sectors. The above expression can be rewritten as: 

$$ d \log W_i = d \log GNE_i  - (\mathbf{e}_i \otimes \mathbf{1}_{J})' \bm{\chi} \frac{\partial \log \mathbf{p}}{\partial \mathbf{h}} (\bm{\rho} + d\mathbf{CBAM})
$$

 which, in turn, becomes:

\small{
$$d \log W_i =   \left[  \frac{w_i \overline{L}_i}{I_i} \frac{\partial \log w_i}{\partial \mathbf{h}} + \frac{t_i E_i}{I_i} \left( \mathbf{1}_{\mathcal{N}_{c(i)}} \frac{ \partial \log t_i}{\partial \mathbf{h}} + \mathbf{1}_{\mathcal{N}_{nc(i)}} \frac{ \partial \log E_i}{\partial \mathbf{h}} \right) \right]  (\bm{\rho} + d\mathbf{CBAM}) + $$
$$- (\mathbf{e}_i \otimes \mathbf{1}_{J})' \bm{\chi} \frac{\partial \log \mathbf{p}}{\partial \mathbf{h}} (\bm{\rho} + d\mathbf{CBAM})+ $$}

\normalsize

\begin{equation}
       + (1-\beta) (1-\rho)  \sum_{j \in \mathcal{J}} \lambda_i^j\sum_{r \in \mathcal{J}}   \iota_{ni}^{kr} (\rho +d CBAM_{ni}^{kr}) 
 \end{equation}

 where, given our model assumptions, in equilibrium $\nu_i^j = \chi_i^j$  and $\bm{\chi} = diag(\chi_1^1, ... , \chi_N^J) \in \mathcal{M}(NJ \times NJ)$. $\frac{\partial \log \mathbf{p}}{\partial \mathbf{h}}$ is given by (\ref{derPRICES}), $\mathbf{h} = (h_{ni}^{k1}, h_{ni}^{k2}, ..., h_{ni}^{kN}) \in \mathbb{R}^{NJ} $ and $ \mathbf{CBAM} = (CBAM_{ni}^{k1}, CBAM_{ni}^{k2}, ...,CBAM_{ni}^{kN}) \in \mathbb{R}^{NJ}$, with $dCBAM_{ni}^{kj}$ given by (\ref{derCBAM}). All derivatives are evaluated at the steady state, where $\mathbf{h} =\mathbf{0}$.

\begin{proposition}
\textbf{{Effect on shocks on emissions embodied in imports}}
The first-order approximation around the steady-state of the change in emissions embodied in direct and indirect European imports following the introduction of the CBAM is given by:

$$d \log EEI (\text{EU}) = \sum_{l \in \mathcal{N}} \sum_{s \in \mathcal{N}} \sum_{q \in \mathcal{J}} \sum_{r \in \mathcal{J}} \frac{\partial \log EEI (\text{EU})}{\partial h_{ls}^{qr}} \left( \rho_l^q + d CBAM_{ls}^{qr} \right) =  $$

$$
=  \sum_{l \in \mathcal{N}} \sum_{s \in \mathcal{N}} \sum_{q \in \mathcal{J}} \sum_{r \in \mathcal{J}} \left\{ \left[ \frac{\partial \log diag(\mathbf{e})}{\partial h_{ls}^{qr}} + \bm{\tilde{\rho}}(\mathbf{I}- \gamma\mathbf{\Pi'})^{-1} \bm{\gamma} \frac{\partial \mathbf{\tilde{\Omega}}}{\partial h_{ls}^{qr}}  \bm{\rho}^{-1} + \right. \right. $$

\small{
$$ \left. \left. + \bm{\tilde{\rho}} (\mathbf{I} - \bm{\gamma} \mathbf{\Pi'})^{-1} \bm{\gamma} \frac{\partial \mathbf{\tilde{\Omega}_{EU}}}{\partial h_{ls}^{qr}} diag(\mathbf{\Lambda}) \mathbf{EEI}({\text{EU}})^{-1} \right] \mathbf{v} \mathbf{1}' \right\} \left( \rho_l^q + d CBAM_{ls}^{qr} \right)
$$}
\normalsize

 where $dCBAM_{ls}^{qr}$ is given by (\ref{eqCBAM}) and $\frac{\partial \log \tilde{\omega}_{nm}^{kh}}{\partial h_{ls}^{qr}}$ by (\ref{eqCOSTSHARES}).\\
\end{proposition}

\textit{Proof.} Emissions embodied in trade are calculated using the vector of production-based emissions and input-output multipliers. We borrow the definition from the OECD \parencite{yamano2020co2} and we adapt it to align with the structure of our model. The emissions embodied in imports are given by the following vector $\mathbf{EEI} \in \mathbb{R}^{NJ}$ with entries corresponding to the emissions generated from the production of each good $j \in \mathcal{J}$ in every country $i \in \mathcal{N}$ embodied in imports from all trade partners. In particular, the first equivalence in equation \eqref{eq:EEI} corresponds to the OECD definition, while the subsequent equalities restates it in terms of our notation and modeling framework:

$$ \mathbf{EEI} \equiv \bm{\tilde{\rho}} (\mathbf{I} - \mathbf{\Omega})^{-1} \mathbf{X} \mathbf{1}  = \left[ \bm{\tilde{\rho}} (\mathbf{I} - \mathbf{\Omega})^{-1} \gamma \mathbf{\tilde{\Omega}} diag(\mathbf{\Lambda}) \mathbf{1} \right] GNE $$

 where $\bm{\tilde{\rho}} = diag (\tilde{\rho}_1^1, ..., \tilde{\rho}_N^J) \in \mathcal{M}(NJ \times NJ)$ is a diagonal matrix with entries $\tilde{\rho}_i^j = e_i^j /p_i^j q_i^j \ \forall i \in \mathcal{N}, j \in \mathcal{J}$, $\mathbf{X} \in \mathbb{M}(NJ \times NJ)$ is the matrix of trade flows and we use the fact that each element $p_{ni}^{kj} z_{ni}^{kj} \in \mathbf{X}$ can be rewritten as $p_{ni}^{kj} z_{ni}^{kj} = \gamma_i^j \tilde{\omega}_{ni}^{kj} \lambda_i^j GNE$.
 Given our interest in the emissions embodied in all intermediate goods imported by European producers, we first consider the change in production-related emissions of each country-sector pair embedded in the goods exported to the EU-ETS and then aggregate them across all exporting countries. First, denote by $\mathbf{\tilde{\Omega}}_{EU}$ the matrix of input-output coefficients, with non-zero entries only for European destination sectors, excluding intra-EU trade. Then:

$$ \mathbf{EEI}({\text{EU})} = \left[ \bm{\tilde{\rho}} (\mathbf{I} - \mathbf{\Omega})^{-1} \bm{\gamma} \mathbf{\tilde{\Omega}}_{EU} diag(\mathbf{\Lambda}) \mathbf{1} \right] GNE $$

 And, taking derivatives and evaluating them at the steady-state, we have that the change in embodied emissions in goods imported by the European producers is given by:

$$ \frac{\partial \mathbf{EEI}(\text{EU})}{\partial h_{ls}^{qr}} =  \frac{\partial \bm{\tilde{\rho}}}{\partial h_{ls}^{qr}} \bm{\tilde{\rho}}^{-1} \bm{\tilde{\rho}} (\mathbf{I} - \bm{\gamma} \mathbf{\Pi'})^{-1} \gamma \mathbf{\Pi'}_{EU} diag(\mathbf{\Lambda})\mathbf{1}  + $$

$$ + \bm{\tilde{\rho}}  (\mathbf{I} - \bm{\gamma} \mathbf{\Pi'})^{-1} \frac{\partial (\mathbf{I} - \mathbf{\Omega})}{\partial h_{ls}^{qr}}  (\mathbf{I} - \bm{\gamma} \mathbf{\Pi'})^{-1} \bm{\gamma} \mathbf{\Pi'}_{EU} diag(\mathbf{\Lambda}) \mathbf{1}  + $$

$$ + \bm{\tilde{\rho}}  (\mathbf{I} - \bm{\gamma} \mathbf{\Pi'})^{-1} \bm{\gamma} \frac{\partial \mathbf{\tilde{\Omega}_{EU}}}{\partial h_{ls}^{qr}} diag(\mathbf{\Lambda})  \mathbf{1} + \bm{\tilde{\rho}}  (\mathbf{I} - \bm{\gamma} \mathbf{\Pi'})^{-1} \bm{\gamma} \mathbf{\tilde{\Omega}_{EU}} \frac{\partial diag(\mathbf{\Lambda})}{\partial h_{ls}^{qr}} \mathbf{1}  $$

 where we use the fact that world $GNE$ is normalised to 1. Readjusting terms: 

$$\frac{\partial \log \mathbf{EEI}({\text{EU}})}{\partial h_{ls}^{qr}} = \frac{\partial \log \bm{\tilde{\rho}}}{\partial h_{ls}^{qr}} + \frac{ \log diag(\mathbf{\Lambda})}{\partial h_{ls}^{qr}} + \bm{\tilde{\rho}}  (\mathbf{I} - \bm{\gamma} \mathbf{\Pi'})^{-1} \bm{\gamma} \frac{\partial \mathbf{\tilde{\Omega}}}{\partial h_{ls}^{qr}}  \bm{\tilde{\rho}}^{-1}  + $$

$$+ \bm{\tilde{\rho}} (\mathbf{I} - \bm{\gamma} \mathbf{\Pi'})^{-1} \bm{\gamma} \frac{\partial \mathbf{\tilde{\Omega}_{EU}}}{\partial h_{ls}^{qr}} diag(\mathbf{\Lambda}) \mathbf{EEI}({\text{EU}})^{-1}  $$

 Recalling that $\tilde{\rho}_i^j = e_i^j / (p_i^j q_i^j) = e_i^j / (\lambda_i^j GNE)$, we have $\log \tilde{\rho}_i^j = log e_i^j - \log \lambda_i^j$ and thus, for each country-sector pair we have: 

$$\frac{\partial \log EEI_{i}^j(\text{EU})}{\partial h_{ls}^{qr}} = \frac{\partial \log e_i^j}{\partial h_{ls}^{qr}} + \tilde{\rho}_i^j \sum_{n \in \mathcal{N}} \sum_{k \in \mathcal{J}} \psi_{in}^{jk} \sum_{m \in \mathcal{N}} \sum_{h \in \mathcal{J}} \frac{1}{\tilde{\rho}_m^h} \omega_{nm}^{kh} \frac{\partial \log \tilde{\omega}_{nm}^{kh}}{\partial h_{ls}^{qr}} + $$
\begin{equation}
    + \frac{1}{EEI_{i}^j(\text{EU})} \tilde{\rho}_i^j \sum_{n \in \mathcal{N \setminus \{\text{EU}\}}} \sum_{k \in \mathcal{J}} \psi_{in}^{jk} \sum_{h \in \mathcal{J}} \omega_{n\text{EU}}^{kh} \frac{\partial \log \tilde{\omega}_{n\text{EU}}^{kh}}{\partial h_{ls}^{qr}} \lambda_{\text{EU}}^h
\end{equation}\

 Finally, define $\mathbf{v}$ as the vector with the first J entries equal to 0, while the rest being equal to the corresponding country-sectoral embodied emissions $EEI_i^j(\text{EU})$ as a share of total emissions embodied in European imports. Aggregating across exporting countries, the change in emissions embodied in goods imported by European producers following the marginal increase in trade wedges $h_{ls}^{qr}$ is given by: 

$$\frac{\partial \log EEI({\text{EU}})}{\partial h_{ls}^{qr}} = \sum_{i \in N \setminus \{\text{EU}\}} \sum_{j \in \mathcal{J}} \frac{EEI_i^j (\text{EU})}{EEI(EU)} \frac{\partial \log EEI_i^j(\text{EU})}{\partial h_{ls}^{qr}} = \frac{\partial \log \mathbf{EEI}(\text{EU})}{\partial h_{ls}^{qr}} \mathbf{v} \mathbf{1}'$$

 And to a first-order approximation, the change in emissions embodied in direct and indirect European imports is given by:

$$d \log EEI (\text{EU}) = \sum_{l \in \mathcal{N}} \sum_{s \in \mathcal{N}} \sum_{q \in \mathcal{J}} \sum_{r \in \mathcal{J}} \frac{\partial \log EEI (\text{EU})}{\partial h_{ls}^{qr}} \left( \rho_l^q + d CBAM_{ls}^{qr} \right) =  $$

$$
=  \sum_{l \in \mathcal{N}} \sum_{s \in \mathcal{N}} \sum_{q \in \mathcal{J}} \sum_{r \in \mathcal{J}} \left\{ \left[ \frac{\partial \log diag(\mathbf{e})}{\partial h_{ls}^{qr}} + \bm{\tilde{\rho}}(\mathbf{I}- \gamma\mathbf{\Pi'})^{-1} \bm{\gamma} \frac{\partial \mathbf{\tilde{\Omega}}}{\partial h_{ls}^{qr}}  \bm{\rho}^{-1} + \right. \right. $$

\small{
$$ \left. \left. + \bm{\tilde{\rho}} (\mathbf{I} - \bm{\gamma} \mathbf{\Pi'})^{-1} \bm{\gamma} \frac{\partial \mathbf{\tilde{\Omega}_{EU}}}{\partial h_{ls}^{qr}} diag(\mathbf{\Lambda}) \mathbf{EEI}({\text{EU}})^{-1} \right] \mathbf{v} \mathbf{1}' \right\} \left( \rho_l^q + d CBAM_{ls}^{qr} \right)
$$}
\normalsize

 where $dCBAM_{ls}^{qr}$ is given by (\ref{eqCBAM}) and $\frac{\partial \log \tilde{\omega}_{nm}^{kh}}{\partial h_{ls}^{qr}}$ by (\ref{eqCOSTSHARES}).\\

 In the special case in which a unique country $i$ introduces the CBAM on a specific good $k$ produced in country $n$ and factor shares are equal across countries and sectors, the above expression simplifies in: 

$$
d \log EEI (\text{EU}) = \left[ \frac{\partial \log diag(\mathbf{e})}{\partial \mathbf{h}} + \bm{\tilde{\rho}}(\mathbf{I}- \gamma\mathbf{\Pi'})^{-1} \gamma \frac{\partial \mathbf{\tilde{\Omega}}}{\partial \mathbf{h}}  \bm{\tilde{\rho}}^{-1} \right. + $$

$$
 + \left. \bm{\tilde{\rho}} (\mathbf{I} - \bm{\gamma} \mathbf{\Pi'})^{-1} \bm{\gamma} \frac{\partial \mathbf{\tilde{\Omega}_{EU}}}{\partial \mathbf{h}} diag(\mathbf{\Lambda}) \mathbf{EEI}({\text{EU}})^{-1} \right] \mathbf{v} (\rho + d\mathbf{CBAM})
$$
where we can distinguish the effect driven by the change in the emissions used in production (\textit{technology effect}) and by the change in sourcing patterns (\textit{reallocation effect}). Specifically: 

$$
d \log EEI (\text{EU}) = \underbrace{\frac{\partial \log diag(\mathbf{e})}{\partial \mathbf{h}} \mathbf{v} (\rho + d\mathbf{CBAM})}_{\textit{technology effect}}  + $$

$$
 +\underbrace{ \left[ \bm{\tilde{\rho}}(\mathbf{I}- \gamma\mathbf{\Pi'})^{-1} \gamma \frac{\partial \mathbf{\tilde{\Omega}}}{\partial \mathbf{h}}  \bm{\tilde{\rho}}^{-1} + \bm{\tilde{\rho}} (\mathbf{I} - \bm{\gamma} \mathbf{\Pi'})^{-1} \bm{\gamma} \frac{\partial \mathbf{\tilde{\Omega}_{EU}}}{\partial \mathbf{h}} diag(\mathbf{\Lambda}) \mathbf{EEI}({\text{EU}})^{-1} \right] \mathbf{v} (\rho + d\mathbf{CBAM})}_{\textit{reallocation effect}} 
$$

\newpage
\subsection{Additional Figures}\label{appendix_fig}

\begin{figure}[ht!]
    \centering
    \includegraphics[width=0.8\linewidth]{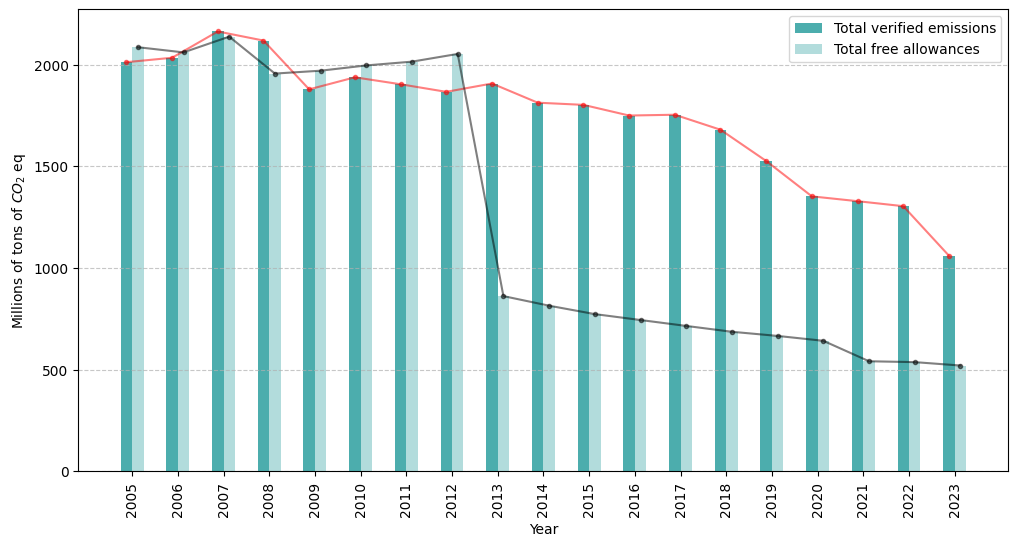}
    \caption{Emissions and Free allowances in the EU ETS (2005-2023) \\
    \textit{Source}: European Union Transaction Log}
    \label{figETSemissions}
\end{figure}
\vspace{1cm}
\begin{figure}[ht!]
    \centering    \includegraphics[width=0.9\linewidth]{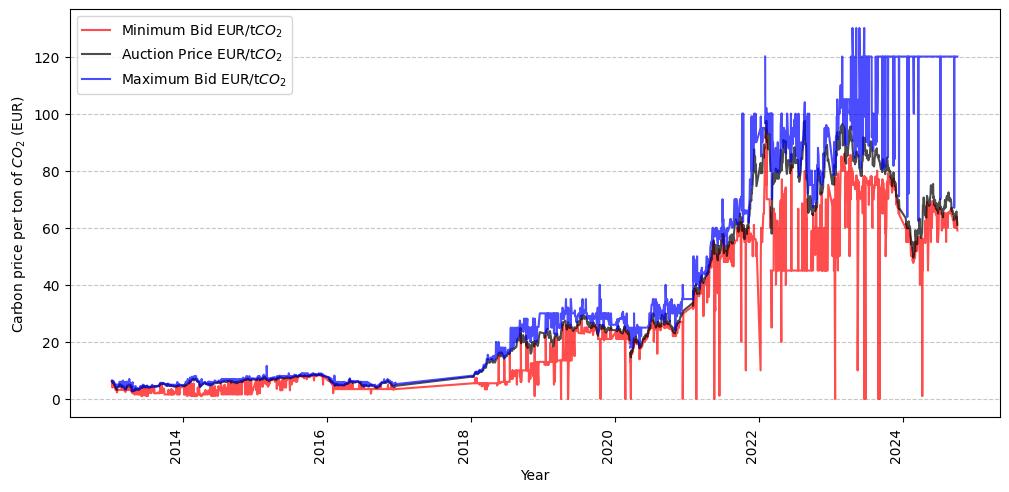}
    \caption{Emissions allowances prices (2013-2024) \\ \textit{Source}: European Energy Exchange}
    \label{figEUAprices}
\end{figure}

\begin{figure}[ht!]
    \centering
    \includegraphics[width=\linewidth]{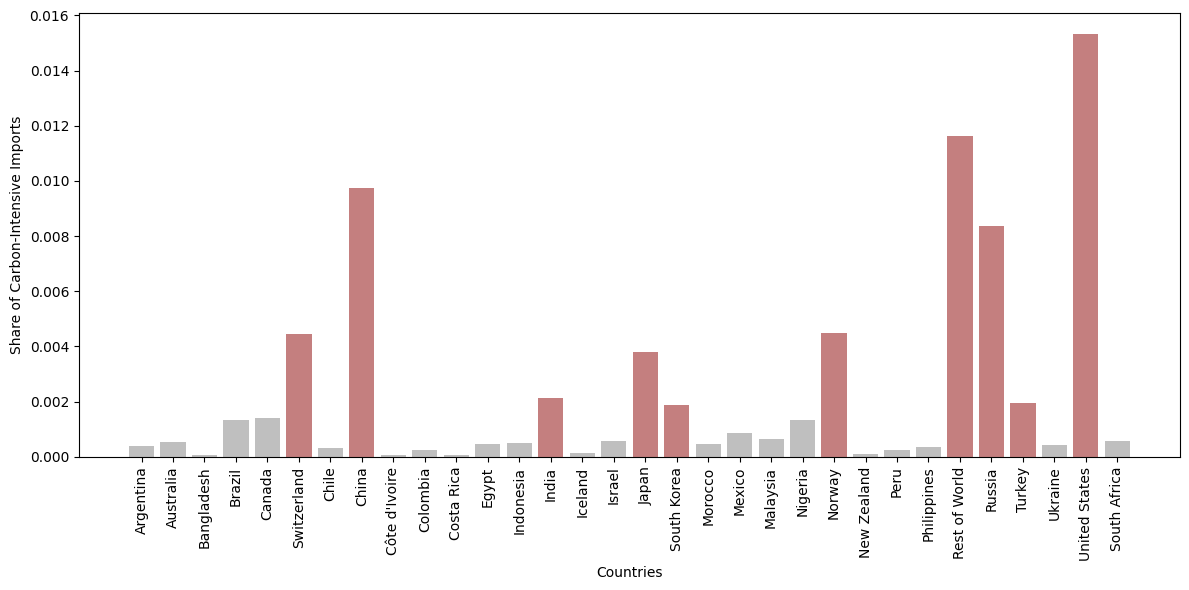}
    \caption{\justifying 2018 Share of EU Carbon Intensive Imports by country in Total Intermediate Purchases. Highlighted, the 10 biggest exporters of carbon-intensive goods to the EU\\
    (in the model, it corresponds to $\tilde{\omega}_{in}^{jk}$ with $n = EU, j \in ETS-sectors$)}
    \label{fig:share_EU_imports_appendix}
\end{figure}

\begin{figure}[ht!]
    \centering
    \includegraphics[width=\linewidth]{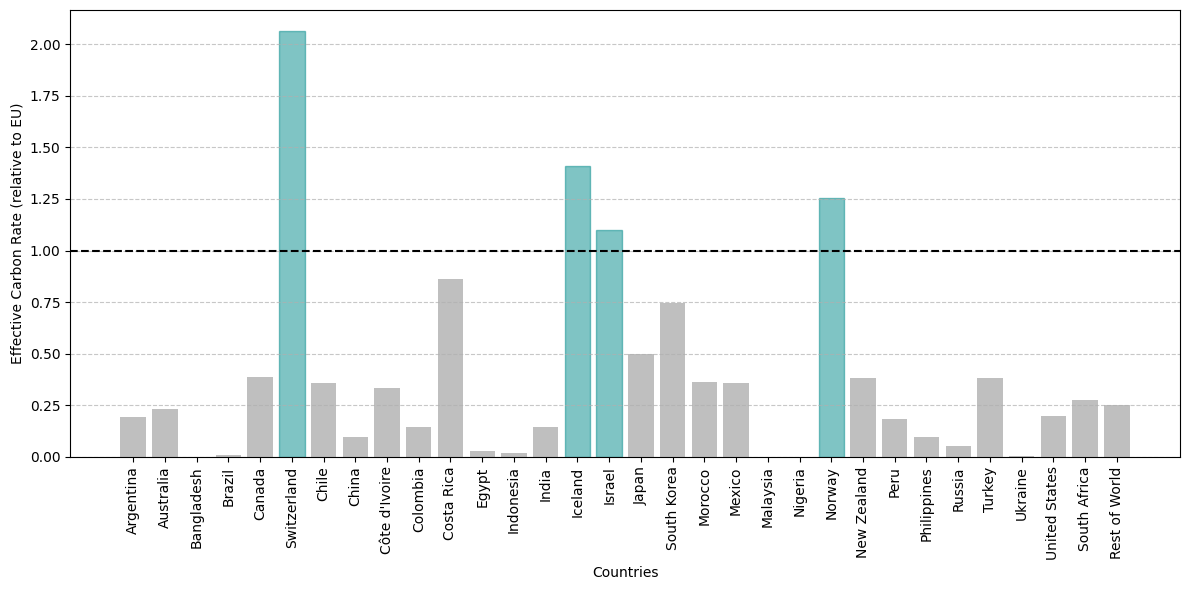}
    \caption{\justifying 2018 Effective Carbon Rates (ECR) by country relative to the European ECR. Highlighted the countries whose ECR is greater than the EU one.}
    \label{figECR_appendix}
\end{figure}

\begin{figure}
\centering
    \includegraphics[width=\textwidth]{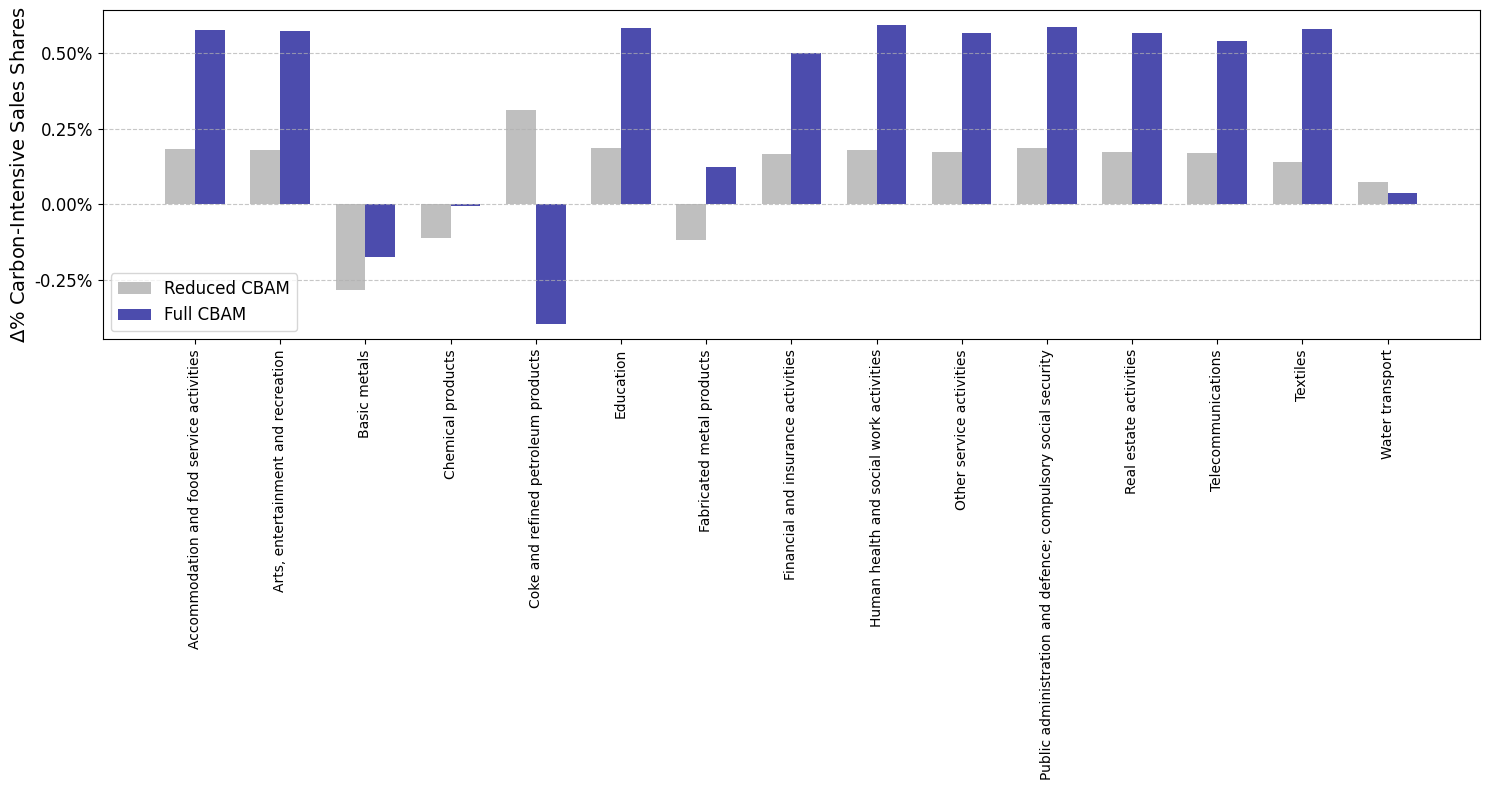}
    \caption{\justifying Policy-induced changes in European Sales Shares of selected sectoral goods (\% of baseline level) -- $\theta = 2$}
    \label{figSALESSHARESCBAM2}
\end{figure}

\begin{figure}
\centering
    \includegraphics[width=\textwidth]{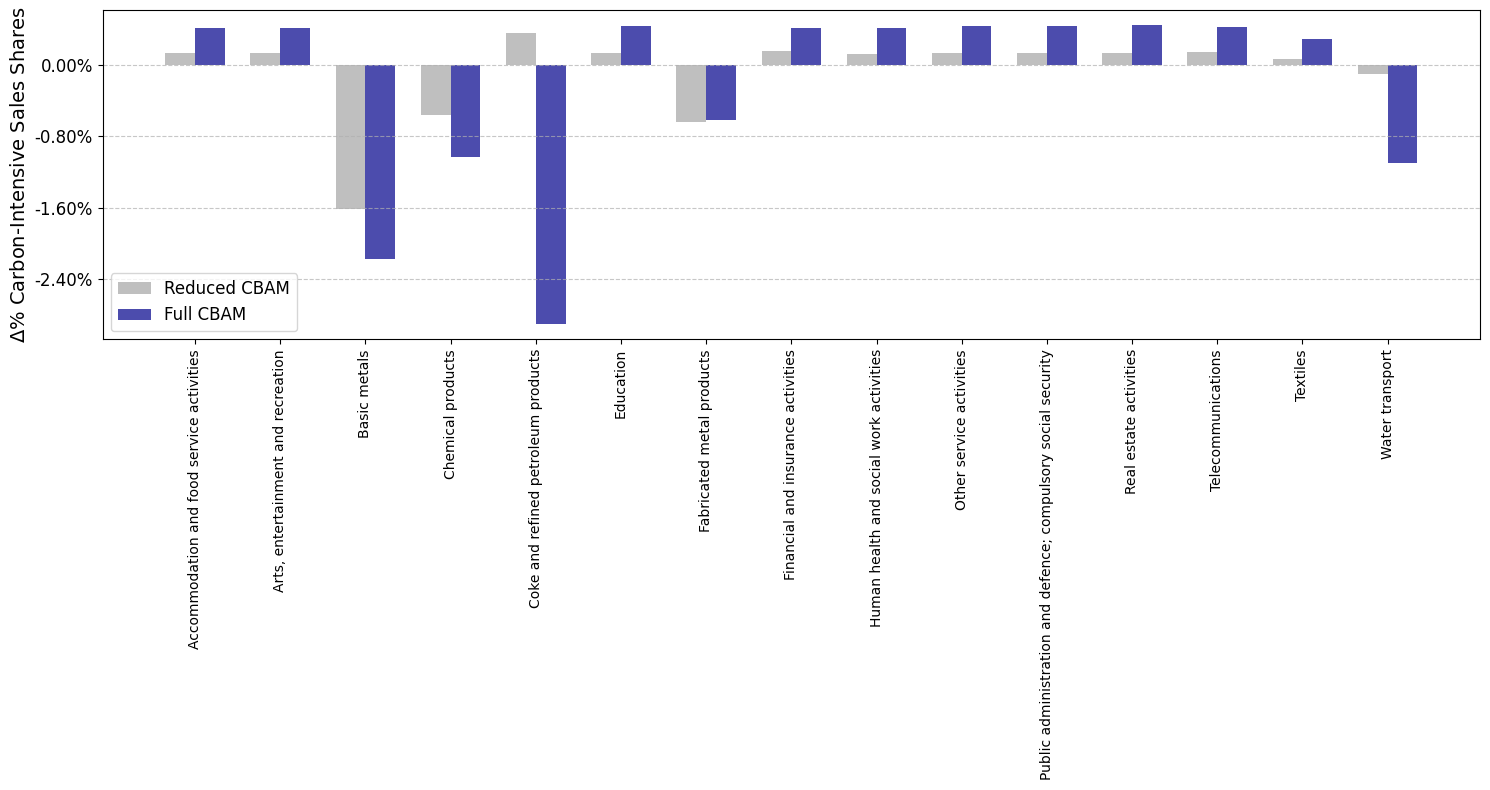}
    \caption{\justifying Changes in EU sales shares of selected sectoral goods (\% of baseline year 2024) -- $\theta = 8$}
    \label{figSALESSHARESCBAM8}
\end{figure}

\clearpage
\subsection{Additional Tables}\label{appendix_tab}

\begin{table}[htb!]
\centering
\scalebox{0.8}{
\begin{tabular}{lrrrrrr}
\hline
Country & Mean & Median & Std & Min & Max \\
\hline
European Union & 0.0082 & 0.0006 & 0.0137 & 0.0001 & 0.0511 \\
Argentina & 0.002 & 0.0005 & 0.0056 & 1.98e-05 & 0.0275 \\
Australia & 0.0042 & 0.0005 & 0.0101 & 5.06e-05 & 0.0506 \\
Bangladesh & 4.50e-08 & 5.89e-09 & 1.60e-07 & 3.29e-10 & 8.23e-07 \\
Brazil & 0.0001 & 3.21e-05 & 0.0003 & 9.29e-07 & 0.0011 \\
Canada & 0.0084 & 0.0021 & 0.0127 & 0.0002 & 0.0616 \\
Switzerland & 0.0048 & 0.0013 & 0.0120 & 0.0002 & 0.0616 \\
Chile & 0.0074 & 0.0017 & 0.0151 & 0.0001 & 0.0660 \\
China & 0.0024 & 0.0004 & 0.0065 & 4.00e-05 & 0.0331 \\
Côte d'Ivoire & 0.0047 & 0.0011 & 0.0113 & 4.96e-05 & 0.0460 \\
Colombia & 0.0018 & 0.0004 & 0.0029 & 9.80e-06 & 0.0120 \\
Costa Rica & 0.0050 & 0.0031 & 0.0068 & 9.23e-05 & 0.031 \\
Egypt & 0.0010 & 0.0002 & 0.0025 & 1.60e-06 & 0.0120 \\
Indonesia & 0.0006 & 0.0001 & 0.0014 & 5.52e-06 & 0.0066 \\
India & 0.0063 & 0.0009 & 0.0175 & 0.0001 & 0.0882 \\
Iceland & 0.0086 & 0.0017 & 0.0180 & 6.63e-05 & 0.0791 \\
Israel & 0.0171 & 0.0011 & 0.0657 & 7.74e-05 & 0.3353 \\
Japan & 0.0078 & 0.0006 & 0.0183 & 2.19e-05 & 0.0839 \\
South Korea & 0.0163 & 0.0012 & 0.0395 & 0.0003 & 0.1505 \\
Morocco & 0.0384 & 0.0022 & 0.1445 & 3.27e-05 & 0.7351 \\
Mexico & 0.0068 & 0.0010 & 0.0214 & 6.96e-05 & 0.1096 \\
Malaysia &  4.40e-08 & 7.34e-09 & 1.37e-07 & 4.08e-10 & 7.00e-07 \\
Nigeria & 4.04e-08 & 2.73e-09 & 8.00e-08 & 4.85e-11 & 3.47e-07 \\
Norway & 0.0062 & 0.0007 & 0.0117 & 3.06e-07 & 0.0400 \\
New Zealand & 0.0030 & 0.0004 & 0.0052 & 3.37e-06 & 0.0162 \\
Peru & 0.0022 & 0.0005 & 0.0049 & 2.90-05 & 0.0192 \\
Philippines & 0.0023 & 0.0002 & 0.0062 & 1.55e-05 & 0.0316 \\
Russia & 0.0022 & 0.0004 & 0.0047 & 3.20e-05 & 0.0222 \\
Turkey & 0.0081 & 0.0010 & 0.0210 & 3.61e-05 & 0.0910 \\
Ukraine & 0.0002 & 2.22e-05 & 0.0003 & 3.67e-06 & 0.0013 \\
United States & 0.0036 & 0.0009 & 0.0097 & 4.65e-05 & 0.050 \\
South Africa & 0.0171 & 0.0015 & 0.0457 & 0.0001 & 0.2310 \\
Rest of the World & 0.0042 & 0.0014 & 0.0090 & 0.0001 & 0.0436 \\
\hline
\end{tabular}}
\caption{\justifying Country-level descriptive statistics of carbon intensities ($\rho_i^j$) in carbon-intensive sectors.}
\label{tab:rhos_appendix}
\end{table}

\renewcommand{\arraystretch}{1.5}

\begin{table}[p]
\centering
\scalebox{0.7}{
\begin{tabular}{l c c c c c c c c}
\toprule
\multirow{3}{*}{\textbf{Variable}} 
  & \multicolumn{4}{c}{\textbf{Reduced CBAM}} 
  & \multicolumn{4}{c}{\textbf{Full CBAM}} \\
  \cmidrule(rl){2-5} \cmidrule(rl){6-9}
 & \multicolumn{2}{c}{\textbf{\textit{Endog.}}} &  \multicolumn{2}{c}{\textit{Exog.}} &  \multicolumn{2}{c}{\textbf{\textit{Endog.}}}&  \multicolumn{2}{c}{\textit{Exog.}}  \\
 \cmidrule(rl){2-3} \cmidrule(rl){4-5}  \cmidrule(rl){6-7} \cmidrule(rl){8-9}
 & Mean & St.dev. & Mean & St.dev. & Mean & St.dev. & Mean & St.dev. \\
\midrule
\multicolumn{9}{c}{\textbf{Panel (a)} $\Delta \%$ in EU Share of Foreign Purchases} \\
\textit{Total} 
  & -0.20 & 0.48 & -1.25 & 2.48 
  & -0.37 & 0.85 & -2.88 & 5.99 \\
\quad \dots \textit{of clean intermediate goods} 
  & 0.23 & 0.14 & 1.06 & 0.61 & 
  0.82 & 0.81 & 3.67 & 0.61 \\
\quad \dots \textit{of dirty intermediate goods} 
  & -0.37 & 0.89 & -2.17 & 4.48 
  & -0.70 & 1.19 & -4.66 & 7.44\\
\midrule

\multicolumn{9}{c}{\textbf{Panel (b)} $\Delta \%$ in EU Share of Domestic Purchases} \\
\textit{Total} 
  & 0.01 & & 0.09 
  & & 0.02 & & 0.24 \\
\quad \dots \textit{of clean intermediate goods} 
  & 0.03 &  & 0.15 
  & & 0.07 & & 0.38 \\
\quad \dots \textit{of dirty intermediate goods} 
    & 0.00 & & 0.07
  & & 0.01 & & 0.20 \\
\midrule  
\multicolumn{9}{c}{\textbf{Panel (c)} $\Delta \%$ in EU Domar Weights} \\
\textit{Total} 
  & 0.09 & 0.11 & -0.05 & 0.34  
  & 0.34 & 0.22 & -0.06 & 0.71  \\
\quad \dots \textit{of clean intermediate goods} 
  & 0.14 & 0.06 & 0.08 & 0.14  
  & 0.45 & 0.18 & 0.15 & 0.47 \\
\quad \dots \textit{of dirty intermediate goods} 
  & 0.06 & 0.12 & -0.12 & 0.39
  & 0.27 & 0.22 & -0.18 & 0.80 \\
\bottomrule
\end{tabular}}
\caption{\justifying Changes in EU-ETS shares (\% baseline level year 2024) for changing elasticities, $\theta =2$ and $\theta=8$. Values are country means; standard deviations in parentheses where applicable.}
\label{tableIMPEXP_elasticities}
\end{table}

\begin{table}
\centering
\scalebox{0.7}{
\begin{tabular}{@{}l@{\hskip 6pt}c@{}c@{\hskip 6pt}c@{}c@{}}
\toprule
\multirow{2}{*}{\textbf{Variable}} 
  & \multicolumn{2}{c}{\textbf{Reduced CBAM}} 
  & \multicolumn{2}{c}{\textbf{Full CBAM}} \\
  \cmidrule(l{6pt}r{0pt}){2-3} \cmidrule(l{6pt}r{0pt}){4-5}
& $\theta=2$ & $\theta=8$ & $\theta=2$ & $\theta=8$  \\
\midrule
\multicolumn{5}{c}{\textbf{Panel (a) }$\Delta \% $ tons emissions embodied in direct imports} \\
\textit{Total} & -1.49 & -9.03 & -2.41 & -18.66 \\
\quad \dots \textit{of clean intermediate goods} & 0.53 & 1.38 & 1.47 & 4.42 \\
\quad \dots \textit{of dirty intermediate goods} & -2.26 & -12.92 & -3.89 & -27.26 \\
\midrule
\multicolumn{5}{c}{\textbf{Panel (b) }$\Delta \% $ tons emissions embodied in direct and indirect imports} \\
\textit{Total} & -0.84 & -5.89 & -1.14 & -11.54 \\
\quad \dots \textit{of clean intermediate goods} & 0.54 & 1.46 & 1.51 & 4.58 \\
\quad \dots \textit{of dirty intermediate goods} & -1.20 & -7.75 & -1.82 & -15.61 \\
\midrule
\multicolumn{5}{c}{\textbf{Panel (c) }$\Delta \% $ Emissions Leakage}\\
\textit{Total} &  -0.04 & -0.16 & -0.13 & -0.29 \\
\bottomrule
\end{tabular}}
\caption{\justifying Changes in emissions embodied in imports (\% of baseline level year 2024) for $\theta = 2$ and $\theta = 8$. Values are country means; standard deviations in parentheses where applicable.}
\label{tableEEI_elasticities}
\end{table}

\end{document}